\documentclass[acmsmall]{acmart}
\usepackage{array}
\usepackage{multirow}
\usepackage{xcolor}
\usepackage{svg}
\usepackage{pdfpages}
\usepackage{threeparttable}
\usepackage{colortbl}
\usepackage{hyperref}
\usepackage{svg}
\usepackage{fontawesome}
\usepackage{tcolorbox}
\usepackage{mdframed}
\usepackage{enumitem}
\usepackage{caption}
\usepackage{subcaption}

\usepackage{blindtext}
\usepackage{capt-of} 
\usepackage{tabularx} 
\usepackage{float}    
\usepackage{multirow}

\definecolor{mygray}{gray}{0.25}
\definecolor{mygray2}{gray}{0.8}
\definecolor{mygray3}{gray}{0.9}
\definecolor{amber}{rgb}{1.0, 0.75, 0.0}
\definecolor{mygreen}{rgb}{0.0, 0.5, 0.0}

\AtBeginDocument{%
  \providecommand\BibTeX{{%
    \normalfont B\kern-0.5em{\scshape i\kern-0.25em b}\kern-0.8em\TeX}}}

\copyrightyear{2022}
\acmYear{2022}
\setcopyright{rightsretained}
\acmConference[CHI '22]{CHI Conference on Human Factors in Computing Systems}{April 29-May 5, 2022}{New Orleans, LA, USA}
\acmBooktitle{CHI Conference on Human Factors in Computing Systems (CHI '22), April 29-May 5, 2022, New Orleans, LA, USA}
\acmDOI{10.1145/3491102.3517742}
\acmISBN{978-1-4503-9157-3/22/04}

  {\list{}{\leftmargin=0.2in\rightmargin=0.2in}\item[]}%
  {\endlist}

\begin{document}

\title[Computational Text Analysis of Child-Welfare Casenotes]{Unpacking Invisible Work Practices, Constraints, and Latent Power Relationships in Child Welfare through Casenote Analysis}

\author{Devansh Saxena}
\affiliation{%
  \institution{Marquette University}
  \streetaddress{Cudahy Hall, 1313 W Wisconsin Avenue}
  \city{Milwaukee}
  \state{WI}
  \postcode{53233}
  \country{USA}}
\email{devansh.saxena@marquette.edu}

\author{Erina Seh-young Moon}
\authornote{\textbf{Both authors contributed equally to this research.} Corresponding author email: devansh.saxena@marquette.edu}
\affiliation{%
  \institution{University of Toronto}
  \streetaddress{140 St. George Street}
  \city{Toronto}
  \state{Ontario}
  \country{Canada}}
 \email{erina.moon@mail.utoronto.ca}

\author{Dahlia Shehata}
\authornotemark[1]
\affiliation{%
  \institution{University of Waterloo}
  \streetaddress{David R. Cheriton School Computer Science}
  \city{Waterloo}
  \state{Ontario}
  \country{Canada}}
  \email{dahlia.shehata@uwaterloo.ca}

\author{Shion Guha}
\affiliation{%
  \institution{University of Toronto}
  \streetaddress{140 St. George Street}
  \city{Toronto}
  \state{Ontario}
  \country{Canada}}
 \email{shion.guha@utoronto.ca}

\renewcommand{\shortauthors}{Devansh Saxena et al.}

\begin{abstract}
Caseworkers are trained to write detailed narratives about families in Child-Welfare (CW) which informs collaborative high-stakes decision-making. Unlike other administrative data, these narratives offer a more credible source of information with respect to workers’ interactions with families as well as underscore the role of systemic factors in decision-making. SIGCHI researchers have emphasized the need to understand human discretion at the street-level to be able to design human-centered algorithms for the public sector. In this study, we conducted computational text analysis of casenotes at a child-welfare agency in the midwestern United States and highlight patterns of invisible street-level discretionary work and latent power structures that have direct implications for algorithm design. Casenotes offer a unique lens for policymakers and CW leadership towards understanding the experiences of on-the-ground caseworkers. As a result of this study, we highlight how street-level discretionary work needs to be supported by sociotechnical systems developed through worker-centered design. This study offers the first computational inspection of casenotes and introduces them to the SIGCHI community as a critical data source for studying sociotechnical systems.
\end{abstract}

\ccsdesc[500]{Human-centered computing~Human-computer interaction (HCI)}
\ccsdesc[300]{Human-centered computing~Empirical studies in HCI}
\ccsdesc[100]{Applied computing~Computing in government}

\keywords{algorithmic decision-making, discretion, bureaucracy, child-welfare system}

\maketitle
\vspace{-0.2cm}
\section{Introduction}
\vspace{-0.1cm}
Government agencies in the United States have sought to reduce costs and increase efficiencies in public policy and social services delivery by increasingly adopting information communication technologies (ICTs) \cite{fernandez2017managing, vigoda2011change, eubanks2018automating} that aim to minimize repeated data collection and bureaucratic overhead, provide targeted client services, and improve decision-making processes \cite{legreid2017transcending}. These ICTs have helped public entities continually collect comprehensive cross-sector data including, structured data (e.g., quantitative assessments), unstructured data (e.g., case narratives), and metadata on different attributes of citizens’ interactions with public services \cite{mergel2016big}. Academics, practitioners, and policymakers have used this data to develop algorithmic systems that purportedly lead to more consistent, objective, and defensible decision-making on critical matters related to human lives \cite{eubanks2018automating, saxena2020human, bullock2018sector}. Various public sector services now use algorithms, such as in child-welfare \cite{saxena2020human}, criminal justice \cite{grgic2019human}, job placement \cite{ammitzboll2021street}, and public education \cite{robertson2020if}, often in the form of risk assessments to preemptively recognize and mitigate "risk" \cite{andrejevic2019automating}.

The U.S. Child-Welfare System (CWS) faces significant challenges. CWS has limited resources, burdensome workloads, and high staff turnover \cite{carnochan2013achieving, saxena2020human}, and faces intense public scrutiny on harm caused to children who are removed from their parents \cite{camasso2013decision} but also when child abuse tragedies occur \cite{gajanan_2020}. These challenges have mounted pressure on CWS to employ algorithmic systems and prove that they follow consistent and objective decision-making processes. SIGCHI researchers have made significant contributions in developing algorithms that aid frontline caseworkers in deciding which calls (i.e., allegations of abuse) should be screened in for an investigation \cite{de2020case, chouldechova2018case}. SIGCHI researchers have also used crowdsourcing platforms such as Amazon Mechanical Turk (MTurk) to study people's perceptions of algorithmic decisions and their impact on human judgment \cite{lee2018understanding, green2020algorithmic}. However, as highlighted by recent ethnographic work in CWS \cite{saxena2021framework2, brown2019toward}, there are drawbacks in these studies that need redressing: \textbf{1)} algorithms built from quantitative administrative data in CWS only account for a narrow set of predictors, offering a deficit-based framing of families \cite{saxena2020human}, and \textbf{2)} experiments conducted on crowdsourcing platforms do not account for organizational/legislative constraints or day-to-day bureaucratic protocols that impact decision-making for all cases \cite{saxena2021framework2}. In light of these concerns, SIGCHI researchers have suggested that collaboratively curated caseworker documentation (i.e., caseworkers' narratives) may offer a more holistic picture of street-level interactions and bureaucratic complexities \cite{ammitzboll2021street, saxena2021framework2}. Unlike administrative quantitative data, caseworker narratives offer a more credible source of information by revealing workers' interactions with families, uncertainties in a case, and impact of bureaucratic constraints on decision-making. These narratives offer much of the desiderata necessary for computational narrative analysis \cite{antoniak2019narrative}. Case notes about families are highly contextual but also share core similarities because they describe similar pathways that most families follow in CWS \cite{jedwab2018caseworkers}. For this study, we pose the following research questions --

\vspace{0.1cm}
\begin {itemize} [leftmargin=*]
  \item \small{\textcolor{mygray}{\textbf{RQ1:} \textit{How can computational text analysis help uncover invisible patterns of caseworker labor?}}}
  \item \small{\textcolor{mygray}{\textbf{RQ2:} \textit{How does computational text analysis highlight the constraints placed on caseworker discretion?}}}
  \item \small{\textcolor{mygray}{\textbf{RQ3:} \textit{How can computational text analysis help uncover latent power structures in CWS?}}}
\end{itemize}
\vspace{0.1cm}

To answer these questions, we conducted computational text analysis of casenotes using topic modeling \cite{wallach2006topic}. For \textbf{RQ1}, we analyzed dominant topics over time and uncover patterns of invisible labor conducted by caseworkers. For \textbf{RQ2}, we divided families into three groups based on their number of interactions with CWS and highlight that families in different groups have varying needs. For \textbf{RQ3}, we conducted computational power analysis of the casenotes to uncover latent power structures in CWS. This paper makes the below unique research contributions --

\begin {itemize} [leftmargin=*]
  \item We offer the first computational investigation of child-welfare casenotes and introduce them as an important and useful data source for studying complex sociotechnical systems. 
 
  \item We highlight invisible patterns of street-level work that caseworkers do within the gaps of legislation (and beyond job duties). These patterns were not uncovered in prior ethnographic work at the same CW agency suggesting case narratives can provide rich contextual information.
  
  \item We show how caseworkers navigate different constraints (systemic, temporal, algorithmic, resource etc.) for different needs of families over the life of a case which uncovers nuances and implications for worker-centered technology design beyond algorithmic interventions. 
  
  \item We found how power relationships for key personas in CW (i.e., CW staff, foster parents, birth parent, etc.) change for different family types, complicating the popular narrative of CW workers having the most power in CW cases.  
  
  
  
  
\end{itemize}

We find support that computational text analysis of casenotes can be a powerful tool for developing holistic decision-support tools instead of the popular administrative data-centered risk assessment tools \cite{chouldechova2018case, de2020case} that have been found to be biased \cite{saxena2021framework2}. This answers calls in prior SIGCHI research about the possibility of using case narratives as an important research tool \cite{saxena2020human, ammitzboll2021street}. We advocate combining computational analysis with qualitative explorations to critique sociotechnical systems. Multiple methodological lenses on the same phenomenon will likely provide holistic insights that any single approach may not \cite{muller2016machine, baumer2017comparing}. In the following sections, we first present the current public sector and computational text analysis research within SIGCHI. Next, we discuss our methodology for answering each of the research questions.

\section{Related Work}

\subsection{Public Sector Research within SIGCHI }
The SIGCHI community has been at the forefront of research on how sociotechnical systems are developed and employed within the public sector. The work has been wide-ranging, including studies that examine issues of civic engagement \cite{dow2018between, golsteijn2016sens}, shaping emergent technologies for collaborative work \cite{bodker2017tying, meng2019collaborative}, designing systems centered on participation and empowerment of affected communities \cite{brown2019toward, dombrowski2014government}, and expanding HCI methods for support labor \cite{fox2020worker}. Through the continued employment of digital technologies in the public sector, researchers have also studied how these systems impacted the decision-making latitude of street-level bureaucrats\footnote{A street-level bureaucrat is a professional service worker (e.g., social worker, police officer, teacher) who operates in the frontline of public service provision. They interact closely with clients and make decisions about them based on how they interpret policies relating to the situations at hand \cite{lipsky2010street}.} who traditionally exercised significant autonomy when implementing policies \cite{busch2018digital}. Studies have found that value conflicts arise when the logics embedded within government's digital platforms do not align with street-level bureaucrats' discretion when they tried enacting the same shared values in practice \cite{voida2014shared, dombrowski2014government, holten2020shifting}. SIGCHI researchers have also unpacked the forms, limits, and complexities of participatory design within the public sector that is now increasingly dictated by public-private partnerships \cite{dow2018between, lodato2018institutional} as newer technologies are designed for the governance of smart cities \cite{whitney2021hci, shadowen2020participatory, heitlinger2019right}.

The continued employment of digital technologies in the public sector has changed governance practices in two distinct ways. First, these systems have improved data sharing practices between different government sectors and purportedly allowed for minimal repeated information gathering, provided targeted services to clients, and allowed for \textit{end-to-end} service delivery \cite{fernandez2017managing, vigoda2011change, eubanks2018automating}. This has allowed government agencies to continually collect data about citizens during their daily operations \cite{mergel2016big}, with the expectation that the data will be transformed into knowledge to inform future decisions that seek to efficiently allocate resources \cite{holten2020shifting}. Here, \textbf{"data becomes the promise of future bureaucratic efficiencies"} \cite{holten2020shifting}. Second, with a primary focus on efficiency and economy, scholars are questioning the core nature of public services as "caring platforms" designed for the public good as opposed to private corporate entities who focus more on optimizing profits \cite{light2019breakdown}. That is, public services that exist to "care for" and "serve" citizens cannot and should not be optimized using performance metrics of the corporate world. SIGCHI scholars have thus begun studying data-driven practices that adopt \textit{care} as a design lens to create systems that advocate for a caring democracy \cite{meng2019collaborative, toombs2018designing, golsteijn2016sens}. Despite two decades of adoption of digital technologies (often referred to as Digital Era Governance \cite{busch2018digital}) and promises of transformation, these tools have generally fused onto existing human discretionary practices rather than altering them at a deeper organizational level \cite{veale2018fairness, saxena2021framework2, holten2020shifting}. Digital technologies have raised the need to understand human discretionary work conducted by bureaucrats who must balance citizens' needs  against the demands of policymakers as they acquire new skills and learn to make decisions through these systems \cite{saxena2021framework2}.

As a result, recent HCI scholarly work has sought to unpack the nature of human discretionary work conducted at the street-level in public services \cite{alkhatib2019street, paakkonen2020bureaucracy, saxena2021framework2, holten2020shifting, robertson2021modeling, clancy2022reconciling}. Alkhatib and Bernstein introduced the theory of \textit{street-level algorithms} to distinctly highlight the gaps in algorithmic decision-making that human discretion needed to address \cite{alkhatib2019street}. Unlike street-level bureaucrats who used discretion to reflexively make decisions about novel cases, street-level algorithms produced illogical decisions that could only be redressed in the future through new data. Pääkkönen et al. further developed this theory to highlight that algorithm design must identify and cultivate important sources of uncertainty because it was at these locations that human discretionary work was most needed \cite{paakkonen2020bureaucracy}. Recently, Saxena et al. \cite{saxena2021framework2} synthesized this prior work conducted in the public sector into a cohesive framework for algorithmic decision-making adapted for the public sector (ADMAPS) which accounts for and balances the complex interdependencies between human discretion, algorithmic decision-making, and bureaucratic processes. ADMAPS framework encourages developing algorithms based on a holistic decision-making process, balancing complex dynamics within sociotechnical systems, and accounting for human discretion and bureaucratic processes \cite{saxena2021framework2}.  Additionally, Ammitzbøll Flügge et al. \cite{ammitzboll2021street} and Saxena et al. \cite{saxena2021framework2} highlighted the collaborative nature of caseworkers' decision-making processes and the impact of bureaucratic structures that algorithm design need to account for. In sum, HCI scholars have reached a general consensus that any algorithmic interventions in the public sector needed to understand the complexities of human discretion carried out at the street-level when implementing day-to-day bureaucratic processes and legislative policies.

\subsection{Child Welfare Research within SIGCHI}
Recent work within SIGCHI has focused on understanding how we can better support individuals and groups within CW. Gray at al. \cite{gray2019trove} have worked on designing technologies for foster youth by creating a new digital memory box for fostered and adopted children to create and store their childhood memories. Badillo-Urquiola et al. \cite{badillo2018chibest} have focused on addressing online safety within foster families by identifying the challenges foster parents face as they mediate teen technology use in the home. Recently, the community has expanded its efforts towards understanding algorithmic decision-making systems employed in CWS \cite{saxena2021framework2, de2020case, cheng2021soliciting}. Algorithms are currently used to determine if a child should be removed from a parent’s care \cite{cuccaro2017risk}, the level of care a child needs \cite{moore2016assessing}, and the type and intensity of services a family will receive \cite{bosk2018counts}. While these decisions can be life-altering, a systematic review of CWS algorithms has shown that many failed to incorporate child-welfare literature or social science theories, instead primarily adopting a deficit-based framework that performed poorly against outliers and deviated from target outcomes \cite{saxena2020human}. SIGCHI researchers have also directly engaged with CWS stakeholders (i.e., families, frontline workers, and specialists) to understand community perspectives and the impact of algorithms on frontline workers’ decisions. Brown et al. \cite{brown2019toward} conducted community-based co-design workshops with CWS stakeholders and found that they felt uncomfortable with algorithmic systems because decisions were centered in deficit-based frameworks that perpetuated biases and bolstered distrust. Complementing this work, De-Arteaga et al. \cite{de2020case} found that frontline workers sought supervisor approval to override an algorithmic decision when they considered it to be incorrect. Similarly, Cheng et al. \cite{cheng2021soliciting} examined stakeholders' understanding of `fairness' regarding machine learning systems in CWS and proposed a framework that allows stakeholders' notions of fairness to emerge organically by working directly with public sector agencies to develop systems that provide a higher comfort level to the community.

Recent ethnographic work in CWS also revealed caseworkers’ frustrations with state-mandated algorithms as they did not account for an agency's resource constraints, legislative policies, or uncertainties inherently present in every case \cite{saxena2021framework2}. Saxena et al. \cite{saxena2021framework2} also found that all the caseworkers involved at the front-end of case planning collaboratively curated casenotes comprising details about interactions with families, uncertainties about the case, critical decisions, and sequence of events that offer a more holistic perspective of case circumstances. Prior work in CW has conducted qualitative exploration of case narratives for a small corpus of text to understand the experiences of both mothers and fathers \cite{eastman2019content, mcconnell2017caring}. Our study sought to understand whether it is feasible to use computational text analysis of narratives to uncover critical details about CW cases such as patterns of human discretionary work conducted by caseworkers and the bureaucratic processes that constrain human discretion.

\vspace{-0.2cm}
\subsection{Computational Text Analysis in SIGCHI Research}
The study of sociotechnical systems requires an understanding of how nuanced and contextualized activities of humans inform, shape, and are shaped by technical systems \cite{ackerman2000}. Studying sociotechnical systems often involves the analysis of text data to understand these types of interactions. While scholars often used qualitative methods to analyze such texts in the past, researchers such as Muller et al. \cite{muller2016machine} have found parallels between qualitative methods and machine learning (ML) techniques and explored the possibility of adopting computational text analysis for unstructured text-based datasets. Recently, computational text analysis methods, including ML methods, have become popular in studying sociotechnical systems in the SIGCHI community \cite{ birthstories2019, MIS2016, regrets2018, thighgap2016}. This is because, as Molina and Garip \cite{mlforsociology} note, ML techniques can overcome the long-standing limitations of statistical modeling and provide contextual findings. Moreover, Nguyen et al. \cite{howwedo2020} state that applying computational text analysis on text which is inherently steeped in cultural and social factors can scale to large bodies of text, help discover insights that may only reveal themselves when text is aggregated, unpack subtle patterns, and detect sentiment.

While SIGCHI has widely adopted computational text analysis methods to study sociotechnical
systems, few studies have examined complex sociotechnical systems in the public sector. Instead, much SIGCHI work has only indirectly touched upon areas of relevance in the public sector using computational text analysis. For example, Antoniak et al. \cite{birthstories2019} studied the experiences of pregnant women via Reddit posts, Chancellor et al. \cite{MIS2016} predicted mental illness severity from Instagram tags, and Guha et al. \cite{regrets2018} examined the role of an individual's agency in social media non-use from web survey responses. Of these works, Antoniak et al. \cite{birthstories2019} revealed the versatility and applicability of using computational text analysis on unstructured narrative texts. The authors \cite{birthstories2019} show that topic modeling works well with stories that follow a formulaic sequence of events and can reveal latent power dynamics between personas and patterns of topic transitions. Recently, in the area of sociotechnical systems research in the public sector, Saxena et al. \cite{saxena2020human} conducted a systematic literature review of computational methods used in CWS and suggested employing computational text analysis techniques (e.g., topic modeling) to elicit context-specific information about CWS cases that current statistical and machine learning algorithms fail to draw out.

Our survey of prior literature shows that while much of SIGCHI research has indirectly examined sociotechnical systems in the public sector, there is a dearth of SIGCHI research that employs computational text analysis to examine these complex systems. And yet, outside of the SIGCHI community, scholars have actively examined the utility of applying computational text analysis methods (specifically topic modeling) to sociotechnical systems research in the public sector \cite{tmforqualpolicy, congress2016, smokingban, realityandml2020} and have noted that topic modeling methods can aid qualitative methods by guiding the systematic discovery of information \cite{tmforqualpolicy} and help reduce directionality bias that arises from manual interpretations of text \cite{realityandml2020}. Therefore, responding to these calls by SIGCHI scholars, we employed topic modeling techniques for analyzing child-welfare casenotes. Using topic modeling, we discovered invisible patterns of human discretionary work performed by caseworkers to gain a more holistic understanding of child-welfare work practices with direct implications for algorithmic decision-making and worker-centered systems design.

\vspace{-0.2cm}

\section{Research Context}

\begin{table}[]
\Small
\begin{tabular}{>{\raggedright}p{1.8cm}|>{\raggedright\arraybackslash}p{11cm}}
\hline
\textbf{Heading} & \textbf{Details} \\
\hline
Family Interaction & Describe the frequency/location, quality of interaction, justification for the type and level of interaction (supervised/unsupervised), and conversations with the parent(s)/caregiver(s) regarding what needs to happen in order to move to a lesser restrictive setting of Family Time. \\
\arrayrulecolor{mygray2}\hline
Concerns & Discuss any concern(s) surrounding family time, how they are being addressed, and enter information about future plans to resolve the concern(s)\\
\arrayrulecolor{mygray2}\hline
Communication & Describe the parent’s/caregiver’s response or receptiveness to communicating with the child(ren)’s caregiver(s) and describe any schedule or method of communication.\\
\arrayrulecolor{mygray2}\hline
Special Considerations & Include information on any special considerations for the child and parent(s) during family time (e.g., no contacts orders, parents confirming the visit, anyone who should not be at the visit). \\
\arrayrulecolor{black}\hline
\end{tabular}
\caption{Agency guidelines on how to record visitations in case notes.}
\vspace{-0.8cm}
\label{tab:casenote_example}
\end{table}

We partnered with a child welfare agency that serves around 900 families and 1300 children in a metropolitan area in a U.S. Midwestern state. The state's Department of Children and Families (DCF) has contracted this agency to provide child welfare and family services and must comply with all DCF standards, including the use of mandated decision-making algorithms. DCF's Initial Assessment (IA) workers investigate allegations of child maltreatment, and if abuse/neglect is substantiated, the case is referred to the agency to provide services. These services are negotiated between the parents' attorneys, district attorney's office, and the judge after caseworkers have conducted initial structured assessments and provided their recommendations to the court. As depicted in Table \ref{tab:cw_teams}, this agency is comprised of several different child-welfare teams that work in collaboration based on the specific needs of families. From the onset of a case, a safety and permanency plan is developed which also establishes the frequency of interactions between caseworkers and birth parents, and consequently, the documentation of these interactions. The agency has established rigorous standards around writing casenotes and compiling case documentation since information needs to be shared among all involved parties (i.e., CW staff, parents' attorneys, district attorney's office, judge). Caseworkers are trained at the agency to write detailed, narrative-style casenotes to record information about families based on observations, pertinent details, and discussions with families. The agency's training guide on case notes is informed by best practices in social work literature \cite{cpsmanual, geiger2021assessment}. For instance, Table \ref{tab:casenote_example} provides an example of how caseworkers must record supervised visits in casenotes.  This collaboratively curated documentation by CW staff involved at the front-end of case planning also acts as a roadmap of decisions made (and the circumstances surrounding these decisions) if such decisions need to be critiqued and/or defended for any case. Narratives, unlike risk assessments, also capture the uncertainties inherent in any child-welfare case. Understanding these uncertainties (and their impact on caseworkers' decisions) becomes especially important for cases where a child-welfare tragedy may have occurred. Prior work \cite{callahan2018paradox} in CWS highlighted these uncertainties for a case where a child passed away - 

\begin{quote}
\textit{\textcolor{mygray}{"How can the uncertainties confounding workers be conveyed in such situations: the deep commitment of the mother to do well by her child, the remorse of the father and his agreement with a court order to stay away, the rallying around of family members and friends, the subsequent loss of the father’s job, the worker’s transfer to another caseload, the move of the family to another community, all occurring over time, amidst improvements in the child’s care, and amongst all of the other factors taking place in the lives of the parents, workers, family members and others."}}
\end{quote}

Case management supervisors add another layer of accountability by ensuring that caseworkers are updating casenotes on a bi-weekly basis and providing detailed descriptions. The agency also has specific instructions in the "Case Note Content Guide" on how to record face-to-face interactions, phone calls, court hearings, and visitations. Many of these uncertainties and complexities are highlighted in casenotes, and we expected computational text analysis on these casenotes could reveal nuanced dynamics between caseworkers and families. CWS comprises of several different child-welfare teams (see Table \ref{tab:cw_teams}) and works with families based on varying case circumstances. We specifically analyze casenotes written by the \textit{\textbf{Family Preservation Services (FPS)}} team that works with birth parents in their efforts to achieve reunification with their children. However, every family is assigned a case management team (case manager and supervisor) that also works with the family and FPS and records their interactions in casenotes which are then compiled in case documentation and made available to all involved parties. We obtained Institutional Review Board (IRB) approval from our mid-sized private research university to use casenotes for research. 

\begin{table}[]
\Small
\begin{tabular}{>{\centering}p{1.5cm}|>{\raggedright}p{3.5cm}|>{\raggedright\arraybackslash}p{7.7cm}}
\hline
\textbf{Abbreviated Name} & \textbf{Details} & \textbf{Role} \\
\hline
IIS & Intensive In-home Services & Provides in-home services to both birth and foster parents where the child has high medical needs \\
\arrayrulecolor{mygray2}\hline
HART & Human Anti-Trafficking Response Team & Manages cases where the foster youth is at high-risk for human or sex trafficking \\
\arrayrulecolor{mygray2}\hline
ICWA & Indian Child Welfare Act & Manages cases concerning children from native American tribes \\
\arrayrulecolor{mygray2}\hline
YTA & Youth Transitioning to Adulthood & Work with foster youth who are about to age out of the foster care system and require independent living provisions \\
\arrayrulecolor{mygray2}\hline
FPS & Family Preservation Services & Works with birth parents in their efforts to achieve reunification \\
\arrayrulecolor{mygray2}\hline
FCA & Foster Care and Adoption & Works with foster parents for training and certification to manage children's needs, foster care licensing, and adoption \\
\arrayrulecolor{mygray2}\hline
PC & Permanency Consultation & Works with case management through the legal process of achieving permanency (i.e., reunification, adoption, or guardianship) \\
\arrayrulecolor{black}\hline
\end{tabular}
\caption{Different kinds of child-welfare teams at the agency}
\label{tab:cw_teams}
\vspace{-0.5cm}
\end{table}

\section{Methods}
This section provides details about the dataset and the data cleaning process followed by our data analyses process. For this study, we employ methodology developed by Antoniak et al. \cite{antoniak2019narrative} for computational narrative analysis using topic modeling. The authors \cite{antoniak2019narrative} showed that their methodology work well for corpus of text that follows a specific sequence of events with frequently occurring personas – characteristics that are observed in child welfare casenotes. 

\subsection{Dataset}
This dataset was acquired from Family Preservation Services (FPS); a specialized child-welfare team whose primary goal is to help birth parents achieve reunification with their children. FPS works closely with birth parents through parenting classes and other court-ordered services to ensure that a safe living environment can be achieved for children. FPS must provide substantial evidence to the DA's office and the judge in order to recommend reunification. They accomplish this by recording parents' progress in parenting classes and other services as well as risk factors within the households. Documenting casenotes is an important task for caseworkers because it guides the child welfare staff on the next steps, provides evidence that agency or caseworkers are making reasonable efforts to help children, and serves as a collaborative tool by demonstrating the collective efforts between families and caseworkers \cite{cpsmanual, geiger2021assessment}. FPS works closely with the case management team and other service providers and also has access to their casenotes which are compiled into case documentation. In this regard, casenotes are collaboratively written by CW staff. Our collaborators at the agency shared that CW staff spent about half their time working on documentation and updated casenotes on a bi-weekly basis (per on-boarding training) such that all parties have timely access to information. However, every casenote contains the date and time for all interactions, even if the case note is electronically updated at a later date.

Casenotes contain a rich source of information about a family’s case and include details about caseworkers' interactions with and observations of parties involved in a case (e.g., birth parents, foster parents, relatives, and children). We obtained records of 9719 casenote entries (the ‘dataset’) for 310 families referred to the agency around May 1, 2019, and worked with Family Preservation until October 14, 2020, or were discharged sooner. Families that received services from the agency were assigned a family identification number, and caseworkers entered casenote details whenever a relevant interaction related to the family took place (e.g., phone call, home visit, parenting class, domestic violence class, court hearing, etc.). Specifically, the dataset contained detailed information on when an interaction related to the family occurred, the duration of the interaction, the time the interaction took place, family member names related to the case, detailed narrative texts on what happened during the interactions, and the caseworker names.

\subsubsection{Data Preparation, Cleaning and Anonymization}
As we were interested in tracking the detailed sequence of interactions between families and CWS staff and inferring how interactions changed over time, we collated all narrative casenotes related to each family identification number in chronological order. Next, we extracted text columns and respective family identification numbers from the collated casenotes. All other columns were excluded from our analysis. We cleaned the collated casenotes by removing punctuation and stopwords from the text. We also anonymized all personal information to protect the privacy of the families. The anonymization process was conducted in the following two steps. First, we used the frequently occurring surnames dataset from the 2010 U.S. Census \cite{census2010surname} and Social Security popular baby names dataset \cite{ssbabynamesWI} to remove all first and last names from the casenotes. We, however, did not remove any first or last names that also functioned as common nouns, such as the last names List and Brown. Second, we replaced all numerical-related information in the text with the word \textit{NUM}. Table \ref{tab:statistics} shows the summary corpus statistics after preprocessing the narrative text. Table \ref{tab:statistics} shows that of the 310 collated casenotes, 235 casenotes contain text greater than 1500 words, and the maximum word length of a casenote is over 38,000 words. Figure \ref{fig:word_distribution}, a violin plot depicts a skew in the word length distribution of casenotes after data curation where most casenotes are shorter in length. 

\begin{minipage}{0.5\textwidth}
\begin{table}[H] 
\Small
\centering      
\begin{tabular}{c | c}  
\hline         
\textbf{Metric} & \textbf{Value} \\ [0.5ex] 
\hline
Number of casenotes with more than 1500 words & 235  \\   
Average number of words per casenote & 3,835  \\ 
Number of words in longest casenote & 38,748  \\ 
Number of unique words & 44,407 \\ [1ex]
\hline      
\end{tabular}
\caption{Corpus Statistics} 
\label{tab:statistics}  
\end{table}
\end{minipage}
\hfill
\begin{minipage}{0.5\textwidth}
\begin{figure}[H]
\vspace{-0.2cm}
 \centering
 \includegraphics[width=0.75\textwidth]{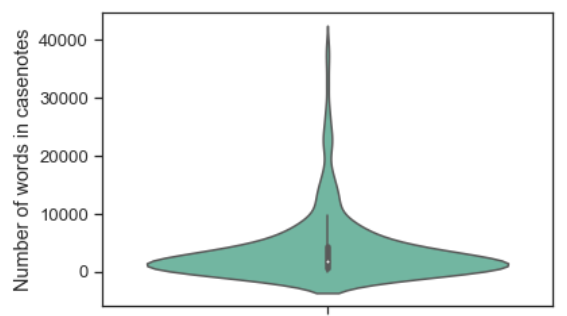} \caption{Word distribution in Casenotes.}
 \label{fig:word_distribution}
\end{figure}
\end{minipage}



\subsection{Topic Modeling and Narrative Analysis Over Time}

\subsubsection{Topic Modeling Solution}\label{section:traintopic}
Topic modeling is one of the most widely used text mining methods in natural language processing (NLP) to infer latent themes from text documents and extract features from bag-of-words representations \cite{ldahanna}. We decided to use LDA for our child-welfare casenotes because LDA can provide easily interpretable insights into densely structured texts which contain both formal and informal language such as ethnographic fields notes \cite{ethno2020} and Reddit stories \cite{birthstories2019}. Following Antoniak et al. \cite{antoniak2019narrative}, we used Mallet’s implementation of LDA topic model to train our topic model. As this implementation of LDA requires the number of topics as a hyperparameter, we took the next two steps to train the topic model. First, we determined the optimal number of topics by creating topic model solutions from 1 to 30 topics and calculated the coherence score and average topic overlap (using the Jaccard similarity statistic) when we assigned 15, 20, 25, and 30 keywords to each of the topics. We found that 14, 17, 22, and 29 topics maximize the divergence between the topic model’s coherence score and average topic overlap. We then manually inspected the 14, 17, 22, and 29 topic model outputs to determine the optimal number of topics. After the interpretations were collaboratively discussed, we reached a consensus to use the 17 topic model solution depicted in Table \ref{tab:six_themes}.

\vspace{-0.3cm}
\subsubsection{Member Checks for Topic Model Qualitative Interpretation}
Topic model outputs often identify thematic patterns in the text at lower abstraction levels than human interpretivist analyses and can benefit from grounded thematic methods to draw out themes in the text \cite{baumer2017comparing}. As such, three co-authors of this paper used an open-coding process on the original casenotes that have the highest probabilities assigned to each topic to capture patterns (themes) within the texts \cite{braunclark2006}. Each co-author individually identified dominant themes, labeled the topics, and then collaboratively discussed their interpretation and labels with co-authors. After this iterative process was complete, a consensus was reached between co-authors on the final trained topic model's themes. Having assigned themes to topics, we next conducted member checks by providing caseworkers with our interpretations of topics, top keywords, and examples of original casenotes with the highest probability (for each of the respective topics). Creswell and Miller \cite{membercheck2010} argue that member checking is crucial to establishing credibility to qualitative analyses as this technique brings study participants back to the data to judge how accurate and realistic researchers' interpretations are. Accordingly, we asked frontline caseworkers to determine the high-level themes based on their reading of the original casenotes and asked if they agreed with our interpretative themes. Caseworkers' feedback helped us further refine our interpretation and topic labels. After these iterative discussions were complete, we reached a consensus about the interpretations of the topics.

\vspace{-0.3cm}
\subsection{Group Analysis of Topic Popularity Over Time}  \label{section:grouping}
Prior work in CWS \cite{maluccio2000child, carnochan2013achieving, tilbury2005counting} has found that caseworkers work with families for different lengths of time depending on the family's unique needs. In addition, CWS experiences a high turnover rate due to high caseloads. To mitigate this phenomenon, CW agencies often group cases in high, medium, and low needs groups based on case severity and assign them to caseworkers to ensure more equitable workloads \cite{kothari2021retention}. Prior work has also highlighted that case complexity (e.g., type of maltreatment, age, number of children, need for financial assistance, drug abuse in the family) is directly associated with the time spent under the care of CWS \cite{pinna2015evidence, carnochan2013achieving}. In line with these studies, we sought to examine if the length of casenotes can serve as a proxy for the family's needs and the severity of the case. To that end, we interrogated the distribution of number of interactions that families have with the child welfare agency. Figure \ref{fig:family_interaction} shows that most families interact with child welfare staff less than 10 times, and there are fewer families that interact with the agency as the number of interactions with the agency increases. Table \ref{tab:summarystat} demonstrates that families in this dataset interacted with CW staff an average of 31 times. Based on the percentile distributions, we grouped families into roughly three equally sized buckets. We then conducted statistical and qualitative analysis into each group's casenotes to determine if the number of interactions with CW staff can serve as a proxy for a family's level of need. Finally, we applied the trained topic model from \ref{section:traintopic} to each group to track topic popularity over time. To accomplish this, we segmented each of the cleaned casenotes into ten equal sections and calculated how average topic probabilities differ for the groups. As casenotes follow a formulaic sequence of events, we were able to divide the texts into ten equal-length sections to create normalized sections (see Fig. 4-8). We define these normalized and chronologically arranged casenote sections as \textbf{"Life of a Case"} which further allowed us to study which topics emerged as significant at different temporal points in a case.

\begin{minipage}{0.5\textwidth}
\begin{table}[H] 
\Small
\centering      
\begin{tabular}{c | c}  
\hline      
\textbf{Descriptive Statistic} & \textbf{Value} \\ [0.5ex] 
\hline
N & 9,616 \\
Mean & 31.1  \\   
Standard deviation & 36.3  \\ 
25 percentile & 7.0 \\
Median & 19.0 \\ 
75 percentile & 40.0 \\ 
\hline     
\end{tabular} 
\vspace{0.5cm}
\caption{Descriptive statistics on family interactions \\ \hspace*{25mm} with CW staff} 
\label{tab:summarystat}  
\end{table}
\end{minipage}
\hfill
\begin{minipage}{0.5\textwidth}
\vspace{-0.2cm}
\begin{figure}[H]
 \centering
 \vspace{-0.5cm}
 \includegraphics[scale=0.35]{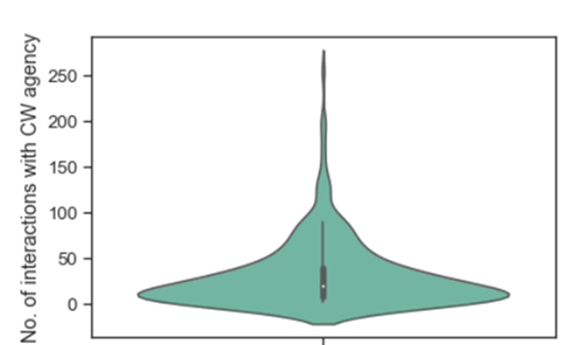}
 \caption{Family Interactions with the agency.}
 \label{fig:family_interaction}
\end{figure}
\end{minipage}


\subsection{Power Analysis of Personas}

\subsubsection{Sentiment analysis}

Child welfare cases involve many parties such as foster parents, family members, and CW staff who are bound by their own responsibilities, goals, and legal obligations. We were interested in examining the power dynamics between such parties by analyzing the day-to-day power relationships between them. However, we needed to first examine the sentiment of casenote sentences because the linguistic choices made by caseworkers could have important implications into how we examine the dynamic relationships between families and caseworkers. 

Caseworkers are trained in writing detailed casenotes based on observations and facts and provide as much descriptive details as possible about their interactions with families \cite{cpsmanual}. As previously noted, this collaboratively curated documentation is imperative for creating a roadmap of critical decisions as well as the circumstances underscoring those decisions. We conducted sentiment analysis on all sentences in the casenotes using a sentiment analysis tool \textit{Valence Aware Dictionary and sentiment Reasoner} (VADER) \cite{vader} to examine the writers' tone of these casenotes. As illustrated by Antoniak et al. \cite{antoniak2019narrative}, VADER was an appropriate tool to compute sentiment analysis since it was developed for social media text and textual data from other domains. Using only sentences with five or more words (to avoid mistakenly segmented sentences and sentences without meaningful information), we assigned a compound sentiment score (a normalized score ranging from -1, extreme negative to +1, extreme positive) to each sentence in the casenotes. As shown in Table \ref{tab:sentiment}, we noted that more than 86\% of the sentences were neutral, and only 9.6\% and 3.5\% of the sentences were classified as positive and negative sentences, respectively. The predominantly neutral tone indicated that the casenotes were mostly descriptive in nature and provided for a suitable corpus of text for conducting power analysis and discovering relationship patterns between key personas.

\begin{table}[ht] 
\Small
\centering      
\begin{tabular}{c | c c}  
\hline          
\textbf{Sentiment} & \textbf{Number of sentences} & \textbf{Percentage} \\ [0.5ex] 
\hline 
\textbf{Positive} & 6,598 & 9.61\% \\
\textbf{Negative} & 2,415 & 3.52\% \\
\textbf{Neutral} & 59,619 & 86.87\% \\[1ex]
\hline     
\end{tabular} 
\caption{Sentiment analysis of casenotes.} 
\label{tab:sentiment}  
\end{table}

\vspace{-0.3cm}
\subsubsection{Personas of Interest}\label{section:replace}
We were interested in examining how power relationships differed between personas in the groups defined in Section \ref{section:grouping}. To do this, we first identified the personas of interest for the whole dataset by manually inspecting the casenotes. Table \ref{tab:personas} illustrates the main personas that appear in all of the casenotes. After identifying personas of interest, we used a non-anonymized version of the casenotes to replace references made to the main personas (References column of Table \ref{tab:personas}) with normalized versions of the persona (Persona column of Table \ref{tab:personas}). For example, we assigned words like \textit{grandmother}, \textit{aunt}, \textit{uncle} to Support System. Table \ref{tab:personas} shows summary statistics on how often these personas appeared in the casenotes, including the total number of mentions of each persona, the number of casenotes that mention the personas, and the average number of times casenotes mention the personas. `Legal parties', `medical parties', and `significant other' rarely appeared in casenotes. As we were interested in measuring the relative power scores between personas, we removed these three infrequently mentioned personas to prevent them from causing potentially statistically spurious effects on power relationship analyses.

\begin{table}[ht!] 
\Small
\centering      
\begin{tabular}{>{\raggedright}p{2cm}|>{\raggedright}p{6cm}|>{\raggedright}p{1.2cm}|>{\raggedright}p{1.2cm}|>{\raggedright\arraybackslash}p{1.8cm}} 
\hline          
\textbf{Persona} & \textbf{References} & \textbf{Total Mentions} & \textbf{Casenotes containing Mentions} & \textbf{Average Mentions per Casenote} \\ [0.5ex] 
\hline 
Biological parent & Mother, Father, Parents, Mom, Dad, \textit{Proper Name} & 29,545 & 281 &	105.14 \\ 
\arrayrulecolor{mygray2}\hline

Child Welfare Staff (CWS) & FPS, OCM, CM, Case manager, Supervisor, FC, \textit{Proper Name} & 17,730 &	277  &	64.01 \\
\arrayrulecolor{mygray2}\hline

Child & Kid, Baby, Son, Daughter, \textit{Proper Name} &	17,935 &	262  &	68.45 \\ 
\arrayrulecolor{mygray2}\hline

Foster parent & Caregiver, FP, \textit{Proper Name} & 3,368 &	148 & 22.76  \\
\arrayrulecolor{mygray2}\hline

Support System  & Grandparents, Aunt, Uncle, MGM, PGM, MGF, PGF, Friend, In-laws, Cousin, \textit{Proper Name} & 1,652 & 125 & 13.22 \\
\arrayrulecolor{mygray2}\hline

Medical Parties & Therapist, Dentist, Doctor, Nurse, \textit{Proper Name} & 334 &	96  &	3.48  \\
\arrayrulecolor{mygray2}\hline

Legal Parties & Lawyer, Judge, Law enforcement, Guardian ad-litem (GAL), Attorney, Assistant district attorney (ADA), District attorney (DA), Court, \textit{Proper Name} &	291 &	89  &	3.27 \\
\arrayrulecolor{mygray2}\hline

Significant Other & SO, Boyfriend, Girlfriend, Significant person, \textit{Proper Name} & 254 &	45  & 5.64 \\ 
\arrayrulecolor{black}\hline
\end{tabular} 
\caption{Persona classification and their most frequent references in text. The \textit{References} column shows the common nouns that are frequently mentioned in the casenotes to represent each persona. Proper nouns (and related variations such as nicknames) are also extracted for the different personas}
\vspace{-0.5cm}
\label{tab:personas}
\end{table}

\begin{table}[!b]
\vspace{-0.2cm}
\Small
\centering
\begin{tabular}{>{\raggedright\arraybackslash}p{12cm}}
\hline           
\textbf{Example Sentences} \\ [0.5ex]  
\hline 
\textcolor{teal}{Sarah [child]} \textbf{demanded} some juice which made the mom upset. \\
This \textcolor{teal}{writer [child welfare staff]} \textbf{communicated} with \textcolor{teal}{Pam [foster parent]} via phone.\\
\textcolor{teal}{Ms. Jones [birth-mom]} \textbf{refused} to speak with \textcolor{brown}{worker [CW staff]} and continuously shrugged her shoulders when asked a question. \\
\hline  
\end{tabular} 
\caption{Paraphrased exemplar sentences depicting power between personas. The child (Sarah) has a high power score; CW staff has equal power with foster parent (Pam); and the birth-mom (Ms. Jones) has a high power score and CW staff has a lower power score.}
\vspace{-0.7cm}
\label{tab:powersentences}
\end{table}

\subsubsection{Power computation}
We adapted the works of Antoniak et al. \cite{birthstories2019} and Sap et al. \cite{sap2017connotation} to compute power scores of and power relationships between personas of interest. Sap et al. \cite{sap2017connotation} created a lexicon of power frames where an entity is assigned a positive power when the entity dominates or exerts a level of control over another entity. This definition of power was appropriate for our study as we anticipated that certain personas would exercise power over other personas in a similar manner. The aforementioned lexicon included 1737 verbs, of which each verb indicated directionality with respect to whom power is assigned. Table \ref{tab:powersentences} shows examples of paraphrased sentences from our casenotes where verbs are assigned power. Next, we lemmatized the verbs in the lexicon and our casenotes, parsed the casenotes which contained the normalized personas from Section \ref{section:replace} using the spaCy parser, and computed power scores for personas of interest in each of the groups by extracting the subjects, verbs, and direct objects from each sentence. Finally, we incremented (or decremented) each persona's power score according to its position in the sentence and the verb power effect. In addition to the results of sentiment analysis, this power analysis method was appropriate here because the goals of all involved personas are aligned and centered in achieving reunification for children and birth parents. 

\section{Results}

\begin{table}[]
\Small
\begin{tabular}{|>{\raggedright}p{0.2cm}|>{\raggedright}p{3.5cm}|>{\raggedright}p{5.3cm}|>{\raggedright\arraybackslash}p{3.5cm}|}
\hline
\textbf{\#} & \textbf{Theme}  & \textbf{Topic} & \textbf{Unique keywords} \\
\hline
\multirow{4}{0.5cm}{1.} & \multirow{4}{3.5cm}{\textbf{Helping Families Secure Resources and Navigate Bureaucratic Processes}} & \textbf{T2}: Helping families secure essential resources & \textcolor{mygray}{\textit{housing, appointment, employment, resources, services}} \\
                    &  & \textbf{T5}: Establishing roles and expectations for different parties & \textcolor{mygray}{\textit{client, reported, shared, meeting, roles}} \\
                    &  & \textbf{T7}: Coordinating virtual interactions during COVID  & \textcolor{mygray}{\textit{virtual, court, camera, communication, covid}}  \\
                    &  & \textbf{T12}: Helping families navigate court proceedings    & \textcolor{mygray}{\textit{court, plan, safety, proceeding, reports}} \\
\hline
\multirow{2}{0.5cm}{2.} & \multirow{2}{3.5cm}{\textbf{Managing Medical Consent, Medication Administration, and Medical Appointments}}    & \textbf{T3}: Managing medical consent between caregivers and accompany clients to medical appointments & \textcolor{mygray}{\textit{caregiver, discussed, consent, form, health}} \\
                    &   & \textbf{T11}: Helping establish medication schedules and manage logistics around medical appointments  & \textcolor{mygray}{\textit{medication, schedule, safety, therapy, appointment}} \\
\hline
\multirow{4}{0.5cm}{3.} & \multirow{4}{3.5cm}{\textbf{Coordinating Time, Travel, and Pickup Logistics for Visitations \& Appointments}} & \textbf{T1}: Managing conflicts between caregivers when scheduling visitations  & \textcolor{mygray}{\textit{visitation, conflict, canceled, voicemail, email}} \\
                    &   & \textbf{T4}: Continued attempts to get in touch with birth parents  &  \textcolor{mygray}{\textit{missed, reschedule, voicemail, phone, contact}} \\
                    &   & \textbf{T6}: Managing logistics around visitations and appointments &  \textcolor{mygray}{\textit{visit, arrived, room, residence, ride, issues}} \\
                    &   & \textbf{T17}: Coordinating travel to and from school for foster children  &  \textcolor{mygray}{\textit{school, attendance, missed, suspended, reports}} \\
\hline
\multirow{2}{0.5cm}{4.} & \multirow{2}{3.5cm}{\textbf{Conducting Structured Assessments to Determine Risks and Progress}} & \textbf{T13}: Keeping track of parents' progress in court-ordered parenting classes  & \textcolor{mygray}{\textit{parenting, chapter, session, curriculum, completed}}\\
                    &   & \textbf{T14}: Conducting home visits, assessing safety concerns, and scoring assessments  &  \textcolor{mygray}{\textit{observed, assessed, home, clean, beds}}\\
\hline
\multirow{2}{0.5cm}{5.} & \multirow{2}{3.5cm}{\textbf{Facilitating Interactions between Children and Parents during Supervised Visits}}  & \textbf{T8}: Observing and facilitating interactions with infants  & \textcolor{mygray}{\textit{baby, visit, bottle, diaper, feeding}}\\
                    &   & \textbf{T9}: Observing and facilitating visits between siblings and adolescents  &  \textcolor{mygray}{\textit{children, play, room, toys, food}}\\
\hline
\multirow{2}{0.5cm}{6.} & \multirow{2}{3.5cm}{\textbf{Observing and Recording Concerns during Transportation}}  & \textbf{T10}: Observing and recording children's behavior during transportation  & \textcolor{mygray}{\textit{transported, slept, cried, picked, visit}}\\
                    &   & \textbf{T15}: Observing and recording pre- and post-transportation concerns & \textcolor{mygray}{\textit{visit, concerns, weather, clothing, seat}}          \\                                           
\hline
\end{tabular}

\caption{17 topic model solution organized by six dominant themes. Topics are labeled T1-T17.}
\vspace{-0.5cm}
\label{tab:six_themes}
\end{table}

\subsection{Topic model solution organized by dominant themes}
We analyzed the results of a topic model solution trained on casenotes and determined 17 to be the optimal topic number based on topic comprehensiveness and interpretability. As illustrated in Table \ref{tab:six_themes}, we further grouped these 17 topics into 6 dominant themes to improve readability. The 17 topics are labeled T1-T17 and all names in exemplar sentences have been replaced with pseudonyms to protect the privacy of individuals.

\subsubsection{\textbf{Helping Families Secure Resources and Navigate Bureaucratic Processes}}
CW staff act as mediators between birth parents, relatives, and foster parents where they help establish roles and expectations for each party as well as bridge the administrative gap between community resource providers, clients, and the court system. 
CW staff work closely with birth parents and help them acquire essential resources that they require to meet their children's needs. They share information with parents about how and where to find resources as well as help them acquire these resources \cite{kellison2019can}. This often takes the form of helping parents find employment, transportation, and home essentials (e.g., food, clothing, toiletries) that would improve stability within the household and facilitate achieving reunification with children. Here, CW staff bridge the gap between community resource providers and clients (i.e., parents) in need. CW staff also work actively to alleviate ambiguity with respect to roles and expectations for each party from the onset of a case \cite{bekaert2021family}. Prior work has established the need to improve communication and enhance teamwork in order to improve relationships in child-welfare practice \cite{bekaert2021family, geiger2017improving}. CW staff also work to ensure that birth parents and foster parents are in agreement with respect to parental roles and expectations. Specifically, CW staff explain to birth parents that foster parents are temporary caregivers who will care for the child's needs and give birth parents the time to make necessary changes within their household so the child can safely return home.

CW staff play a critical role in helping families navigate court proceedings where they escort parents to court and advocate for them. CW staff share progress made by parents in parenting classes, court-ordered services, and parenting skills that they are exhibiting during supervised visits. As illustrated by the exemplar sentence below, CW staff help parents understand the court process and the changes they must make to receive a favorable decision in court. As illustrated by topic 7 (i.e., \textit{virtual interactions during COVID}), CW staff also assumed newer responsibilities during the COVID pandemic in terms of facilitating virtual interactions between parents and children and also helping parents navigate through virtual court hearings. Reading through the case notes, we observed that CW staff also helped parents and caregivers troubleshoot technology issues and explained to them how to use Microsoft Teams or Skype for Business.

\textcolor{mygray}{[T2 Probability: 0.65]} "Case Manager would like the Family Preservation Specialist to visit with Ms. Davidson [birth-mom] at least 1x per week and assist with helping her secure resources, especially for the unborn baby."

\textcolor{mygray}{[T5 Probability: 0.56]} "Family Preservation Specialist [FPS] attended staffing with supervisor, Ongoing Case Manager, and Ongoing Case Manager supervisor to discuss referral, roles, and supportive services needed. FPS attended team meeting to introduce herself to Sarah [birth-mom] and explained her role in the process. FPS asked about what kinds of services Sarah [birth-mom] was in need of and she responded that housing is her main priority. In addition, Family Preservation Services will gather resources on rent assistance, emergency daycare, and baby supplies that Sarah [birth-mom] can then have at her disposal.

\textcolor{mygray}{[T12 Probability: 0.59]} "Family Preservation Services greeted the family and provided them with an introduction of their role and services that they will provide the family. Mr. B [birth-dad] shared that the baby may possibly be placed with his aunt and moving soon. Mr. B [birth-dad] stated that he has court tomorrow at 9am. Family Preservation Services asked the family if they would mind if she attended court with them. Family Preservation Services explained that she would be there for support and to answer any questions that they may possibly have. Mr. B [birth-dad] and Ms. M [birth-mom] agreed and stated that it would be perfect as court can sometimes become confusing."


\subsubsection{\textbf{Managing Medical Consent, Medication Administration, and Medical Appointments}}

Communication between involved parties (i.e., birth parents, foster parents, medical professionals, attorneys) about a child’s medical needs and well-being is essential and is facilitated by CW staff. A foster child may be removed from a parent’s care and placed in temporary protective custody with a foster parent; however, the birth parents still retain their parental rights and decision-making capacity regarding any health services extended to a foster child \cite{browncaring2021}. CW staff work with both the birth parents and foster parents to obtain and manage medical consent such that the foster child can receive medical care in the form of therapy, dental care, or other necessary services. CW staff also help supervise the day-to-day medical needs of foster children by establishing medication schedules as well as accompanying foster children to doctor’s appointments. Here, CW staff's role as a mediator also helps alleviate any conflicts that may arise due to overlapping parenting roles.

\textcolor{mygray}{[T3 Probability: 0.564]} "Family Preservation Specialist met the caregiver, Yvette [foster parent], and Billy [child] at the doctor’s office. Family Preservation Specialist observed Mrs. Olsen [birth-mom] holding Billy [child]. The doctor discussed how Billy [child] was doing and why there were being seen at the clinic. Family Preservation Specialist observed the doctor asking the caregiver questions. Family Preservation Specialist observed Billy [child] have his fists clenched while Mrs. Olsen [birth-mom] held him near the table. Family Preservation Specialist and the caregiver discussed meeting at her home after the appointment."

\textcolor{mygray}{[T11 Probability: 0.68]} "This worker [child welfare staff] attended the case transfer staffing with Ongoing Case Manager in the home of Ms. Brown [birth-mom]. Our group created a medication schedule with Family Preservation Services doing medication observation on Monday's at 7am and Thursday's at 11:00am. Paul [significant other] and Ms. Blar [relative] (maternal aunt to Billy [child]) will observe all other feedings and medication supervisions."


\subsubsection{\textbf{Coordinating Time, Travel, and Pickup Logistics for Visitations and Appointments}}
This theme is centered in the coordination and scheduling work that CW staff undertake in their role as liaisons between birth parents, foster parents, and other professionals in child-welfare. CW staff is responsible for organizing and facilitating supervised visits between children and birth parents \cite{smith2014strengths}. This involves scheduling the time and place of these visits with involved parties, managing scheduling conflicts, as well as transporting children (and parents, if necessary) to the location of supervised visits. These visits may occur at the child-welfare agency, a public space (e.g., public parks), or the parents’ place of residence based on the presence and assessment of impending dangers in the household. CW staff also help parents get access to travel vouchers if they do not have the financial means or a vehicle for traveling long distances. While scheduling, conflicts within a family may also arise. For instance, birth parents might share a contentious relationship and may not want to work with each other. Here, CW staff must also work to ameliorate such concerns in order to promote congruence between involved parties \cite{juul2020collaboration}. This is necessary to ensure that progress is being made towards achieving permanency for the child. In addition, they also help coordinate travel to and from school for foster children who may not have access to a regular school bus route.

\textcolor{mygray}{[T1 Probability: 0.620]} "Family Services Counselor Nadine [child welfare staff] spoke to Mr. Smith [birth-dad] regarding Ms. Smith [birth-mom] and visitation with the children. Mr. Smith [birth-dad] stated that he will not allow Ms. Smith [birth-mom] in his home for visitation."

\textcolor{mygray}{[T4 Probability: 0.949]} "Family Preservation Services contacted Ms. Brow [birth-mom] to schedule a visit with her child and left a message. Family Preservation Services contacted Ms. Brow [birth-mom] and introduced themselves. Ms. Brow [birth-mom] stated that she needed to get off the phone and stated that she would call back. Family Preservation Services called Ms. Brow [birth-mom] to schedule a visit for the next week but Ms. Brow [birth-mom] did not answer the phone. Family Preservation Services left a message."

\textcolor{mygray}{[T6 Probability: 0.598]} Family Preservation Services arrived at Ms. Abel's [foster parent] residence to transport Bob [child] to a supervised visit. Ms. Abel [foster parent] did not report any issues with Bob [child] but she did need to assist with getting him into the vehicle. During the ride Bob [child] was crying but then fell asleep for most of the ride; he did not cause any issues or concerns.


\subsubsection{\textbf{Conducting Structured Assessments to Determine Risks and Progress}}
The child-welfare process is centered on assessing risk factors and helping parents develop protective capabilities to mediate these risks. CW staff, especially Family Preservation Services, works closely with parents through the parenting curriculum (keywords: parenting, chapter, session, curriculum) and other court-ordered services and score their progress on structured assessments. NFCAS (North Carolina Family Assessment Scale) and AAPI (Adult Adolescent Parenting Inventory) are examples of assessments especially used at the child-welfare agency by Family Preservation \cite{kirk2005advances, lawson2017analyzing}. CW staff record a parent’s level of engagement in these classes and whether they are exhibiting changes with respect to how they manage their child’s behaviors. CW staff often refer to this as \textit{perspective shift}, that is, whether the parent understands why their case was referred to CWS and if they are showing the willingness to make necessary changes in their lives. In addition, CW staff conduct home visits to assess safety concerns and any impending dangers within the household. This includes assessing the general cleanliness of the house, availability of food in the pantry and refrigerator, and both the children's and parents’ hygiene. A sanitary and safe home help CW staff implement in-home services such that children do not have to be removed from their home and placed in foster care. Moreover, CW staff are also required to conduct and score quantitative assessments about risk factors associated with parents and children, safety within the household, parents’ life experiences, and parenting skills. As illustrated by Saxena et al. \cite{saxena2021framework2}, these quantitative structured assessments in CWS are now being used to develop algorithmic systems.

\textcolor{mygray}{[T14 Probability: 0.82]} "Family Preservation Services arrived at the home. Ms. Tazan [child welfare staff] was inside with the family. Family Preservation Services and Ms. Tazan did a walk through of the home. The home is not furnished and children don't have beds. Ms. Tazan had all the children clothing in black bags in the closet. Family Preservation Services did not observe any toys, books, etc. Family Preservation Services spoke to Ms. Tazan regarding the initial assessments that he needed to complete with her."

\textcolor{mygray}{[T13 Probability: 0.61]} "Family Preservation Services and Mr. Gibbs [bio-dad] watched the videos together and went through the power point presentation. It was apparent that Mr. Gibbs [bio-dad] had read the material as he was engaged in the discussion and talked about the examples in the book. First parenting assessment completed."


\subsubsection{\textbf{Facilitating Interactions Between Children and Parents During Supervised Visits}}
CW staff, especially Family Preservation Services, help facilitate interactions between children and birth parents and observe how these interactions are going during supervised visits every week \cite{giallo2020preservation}. Family Preservation Services use their expertise in parenting to work with the parents and help improve the quality of these interactions where the parents understand and attend to the needs of their children. Topic 8 (i.e., \textit{interactions with infants}), however, emerged separately as compared to topic 9 (i.e., \textit{interactions between siblings}) because an infant’s interactions (e.g., eating well, sleeping, making eye contact, smiling, etc.) are essentially different from children’s interactions (e.g., playing with siblings, playing with toys, running, etc.) and are noted distinctively by CW staff to assess well-being. For cases where multiple children are involved, CW staff also focus on ensuring that the parent(s) can manage their children's behaviors and establish some disciplinary boundaries. Family Preservation Services works with birth parents and advises them on how to manage interactions between siblings (e.g., fighting, yelling) and how to respond when being challenged by them \cite{giallo2020preservation}. Addressing these concerns helps ensure that time to reunification is reduced and the likelihood of case re-referral is lowered in the future.

\textcolor{mygray}{[T8 Probability: 0.55]} "Ms. Weldon [birth-mom] was excited to see the child as she kissed her and told her how much she missed the child. Ms. Weldon changed the child's clothes and did the child's hair while the child sat in her walker. Ms. Weldon continued to talk about her issues surrounding her case, Family Preservation Services had to remind Ms. Weldon to focus on her daughter instead of her situation she is in. Ms. Weldon praised the child for being able to wave and tried teaching the child how to clap her hands."

\textcolor{mygray}{[T9 Probability: 0.84]} "Ms. Tyndall [birth-mom] met Family Preservation Services outside to help bring in Ned [child], Phil [child], Pete [child], and Lawrence [child] into the family center. Upon entering the family room Ms. Tyndall who was holding Pete's hand and Lawrence in her arms told the boys that they have snacks in her bag for each of them. Ms. Tyndall sat on the floor and let Lawrence crawl and Pete explore in the visit room. Phil and Ned started playing with their little brothers and bringing them toys to play with."


\subsubsection{\textbf{Observing and Recording Concerns During Transportation}}
CW staff are trained to record any issues that may arise before, after, or during transportation \cite{geiger2021foster}. Words in topic 10 (i.e., \textit{children's behavior during transportation}) and topic 15 (i.e., \textit{pre- and post-transportation concerns}) are associated with children’s behavior and/or their interactions with Family Preservation Services while being transported for supervised visits. It helps CW staff assess how to best facilitate a supervised visit. For instance, if a child is anxious and agitated during the drive then CW staff might begin a supervised visit by engaging the child in activities that may help pacify them. This information is also shared and discussed with birth parents and foster parents to assess if there are any traumatic triggers that may be leading to emotional dysregulation. This also involves any concerns that might arise before or after the transportation. For instance, CW staff also ensure that children are dressed appropriately for the weather and look physically healthy.

\textcolor{mygray}{[T10 Probability: 0.48]} "This worker [CW staff] met the family at the Family Center. This worker transported Maya [child] and Jake [child] to their placement in [address]. Maya cried for roughly ten minutes for the car ride and then stopped and played with a stuffed animal. Coordinator Beth [CW staff] asked this worker to inform the caregiver that Maya had cried for roughly one hour during the visitation today. This worker did give this information to the caregiver upon arrival."

\textcolor{mygray}{[T15 Probability: 0.409]} "All three children were transported from maternal grandmother's home located at [address] and transported to McDonalds play land located at [address]. All three children were transported back to grandmother's.  All three children were dressed appropriately for the weather and appeared free of injury, as they were able to walk, run and bend with ease"


\subsection{Group analysis of topic popularity over time}

\begin{figure}[H]
\vspace{-0.3cm}
\centering 
\includegraphics[width=\textwidth]{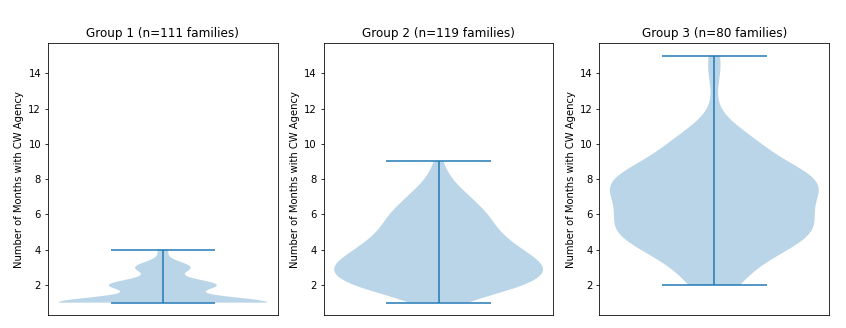}
\caption{Distribution of the frequency of interactions with CW agency for groups G1, G2, and G3}
\label{fig:violin_plot}
\vspace{-0.3cm}
\end{figure}


We divided families into three groups based on their number of interactions with the child-welfare system. Figure \ref{fig:violin_plot} highlights the frequency of interactions that each group had with the agency and the number of months that families in each group worked with the agency. We notice a higher frequency of interactions at the onset of cases because CW staff must follow a 15-month timeline established by the Adoption and Safe Families Act (ASFA) where the State must proceed with the termination of parental rights if reunification has not been achieved in 15 months \cite{adler2001meanings}. Therefore, CW staff work extensively with families from the onset of a case to gather relevant information and take necessary actions to expedite reunification. Below, we further discuss differences among the three groups based on the number of children, birth parents, and foster parents involved in each group. Descriptive characteristics about the three groups are available in Table \ref{tab:group-charts}.

\textbf{Group 1 (G1)} includes \textbf{\textit{Low Needs Families}} that only had 1-10 interactions and generally involve cases of neglect (i.e., lack of childcare, lack of access to healthcare, lack of adequate food or clothing) where birth parents must make necessary changes within their household so as to provide a safe and nurturing environment for their children. As depicted in Table \ref{tab:group-charts} (see Group 1), the majority of the children (n=68, 62\%) were not removed from the care of birth parents, and instead, in-home services were provided to these families. Majority of these families were also single-parent households (n=78, 70\%) and only involved one foster child (n=77, 69\%).

\begin{table}[ht] 
\Small
\centering      
\begin{tabular}{c | c | c c c}  
\hline          
\textbf{Persona} & \textbf{Number} & \textbf{Group 1} & \textbf{Group 2}& \textbf{Group 3}\\ 
& \textbf{of personas} &\textbf{(1 - 10 interactions)} & \textbf{(11 - 40 interactions)}& \textbf{(40+ interactions)}\\ [0.5ex] 
\hline 
 & 0 & 0 & 0 & 0\\
& 1 & 77 & 35 & 20\\
& 2 & 27 & 27 & 24\\
\textbf{Children}& 3 & 3 & 24 & 10\\
& 4 & 2 & 19 & 11\\
& 5 & 2 & 10 & 10\\
& 6 & 0 & 3 & 3\\
& 7 & 0 & 1 & 2\\
\hline 
& 0 & 2 & 0 & 0\\
\textbf{Birth Parents}& 1 & 78 & 70 & 48\\
& 2 & 31 & 49 & 32\\
\hline 
& 0  & 68 & 32 & 19\\
& 1 & 40 & 35 & 50\\
& 2 & 3 & 42 & 6\\
\textbf{Foster Parents} & 3 & 0 & 8& 4 \\
& 4 & 0& 1 & 1\\
& 5 & 0 & 1 & 0\\[1ex]
\hline     
\end{tabular} 
\caption{Descriptive characteristics for the three groups based on the number of interactions with CWS. The table shows the total number of cases for each group $\{1, 2, 3\}$ having $x$ members from the persona list $y$ where $x \in \{[0,7], [0,2], [0,5]\} \text{ for each y} \in \{'Children', 'Birth Parents', 'Foster Parents'\} \text{ respectively.}$ \textit{Zero} value for foster parents means that the child was not removed and in-home services were provided to families.} 
\label{tab:group-charts}  
\vspace{-0.5cm}
\end{table}

\textbf{Group 2 (G2)} includes \textbf{\textit{Medium Needs Families}} that had 11-40 interactions with the child-welfare system. This group includes cases where most children were removed from the care of birth parents and placed with foster parents (or relatives) due to safety concerns within the household. This is generally considered short-term foster care, where birth parents must complete parenting classes and court-ordered services (e.g., drug and alcohol services, domestic violence classes, etc.) and demonstrate stability within the household to achieve reunification with their children. Here, children are generally placed in short-term placements before long-term caregivers can be found. These cases generally involve multiple children placed with different foster parents since it is hard to find foster homes that can provide for all the children involved in a case. As depicted in Table \ref{tab:group-charts}, this group had 35 families with only one child, 27 families with two children, 24 families with three children, and so on. Group 2 also has 32 families where children were not removed, 35 families where children were placed with one foster parent, 42 families where two foster parents were involved, and 8 families where children were split between and placed with 3 foster parents.

\textbf{Group 3 (G3)} includes \textbf{\textit{High Needs Families}} that had 40+ interactions with the child-welfare system and includes cases of more severe abuse and/or neglect. This group is generally considered long-term foster care, where children are placed with long-term caregivers who are trained and certified to care for high-needs children. Foster parents in this group may also be the next of kin since CW staff prioritize placing children with relatives. Prior work has established that children are more likely to achieve emotional and cognitive well-being when placed within the family \cite{denby2017protective, jedwab2020kinship}. However, if children are placed in kinship care, the caregivers assume the role of foster parents (as active caretakers) and are no longer classified as a parent's support system (passive and occasional caretakers). CW staff work closely with birth parent(s) in parenting classes and other court-ordered services as well as help them find stable employment and other resources necessary to meet the needs of their children and eventually achieve reunification. As depicted in Table \ref{tab:group-charts},  this group consists of 19 families where children were not removed from the care of birth parents and in-home services were provided, 50 families where children were placed with one foster parent, 6 families where children were placed with two foster parents and so on. Moreover, similar to Group 2, families in this group consist of multiple children, which adds to the complexity of these cases. 

Next, we discuss trends in topic popularity over time for the top four themes from Section 5.1 for each of the three groups. Following Antoniak et al. \cite{antoniak2019narrative}, we divided each of the casenotes into ten equal sections. As casenotes follow a formulaic sequence of events, we were able to divide the texts into ten chronologically arranged normalized sections (i.e., \textbf{Life of a Case}). This allowed us to track casenotes of varying lengths which begin and end at different times. Therefore, as depicted in Figures 4-8, 10\% on the x-axis would point towards the events happening in 0-10\% of the life of a case; 50\% on the x-axis would point towards events happening in 40\%-50\% of the life of a case.

\subsubsection{\textbf{Helping Families Secure Resources and Navigate Bureaucratic Processes}}
\begin{figure}[H]
\vspace{-0.2cm}
\centering
\begin{subfigure}[b]{0.48\textwidth}
    \centering
    \includegraphics[width=\textwidth]{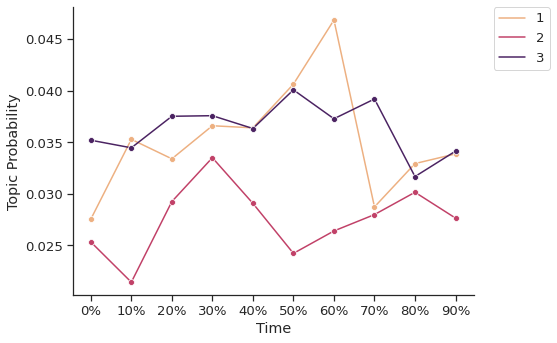}
    \caption[]%
    {{\small \textcolor{mygray}{T2: Securing Resources}}}
\end{subfigure}
\hfill
\begin{subfigure}[b]{0.48\textwidth}  
    \centering 
    \includegraphics[width=\textwidth]{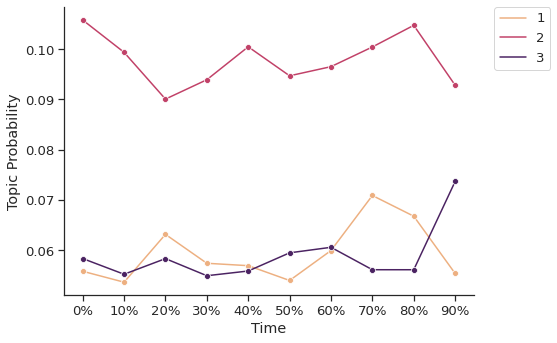}
    \caption[]%
    {{\small \textcolor{mygray}{T5: Establishing roles between parties}}}
\end{subfigure}
\hfill
\vspace{-0.3cm}
\caption{\textcolor{mygray}{Time trends for topics focusing on helping families secure resources and role coordination}}
\label{fig:time_plots_all_2_5}
\vspace{-0.25cm}
\end{figure}

Securing resources (topic 2) is a significant topic for both G1 and G3 families. For G1, we observe an upward trend through the life of a case as shown in Figure \ref{fig:time_plots_all_2_5}(a). CW staff work with birth parents from the onset of a case to acquire these resources to achieve a safe living environment. For G3, CW staff continually work with parents to ensure necessary changes are being made in the household from both an economic and behavioral perspective. However, this topic is less significant for G2 because the more dominant concerns are related to managing logistics (since G2 families involve multiple foster children). Specifically for G2 families (see Figure \ref{fig:time_plots_all_2_5}(b)), CW staff work on managing roles and expectations between birth parents and multiple foster parents as conflicts arise due to overlapping parental roles in managing the needs of multiple foster children.

\begin{figure}[H]
\centering
\begin{subfigure}[b]{0.48\textwidth}  
    \centering 
    \includegraphics[width=\textwidth]{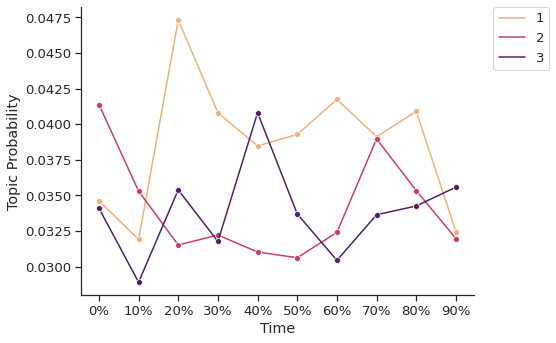}
    \caption[]%
    {{\small \textcolor{mygray}{T12: Navigating Court Proceedings.}}}
\end{subfigure}
\hfill
\begin{subfigure}[b]{0.48\textwidth}
    \centering
    \includegraphics[width=\textwidth]{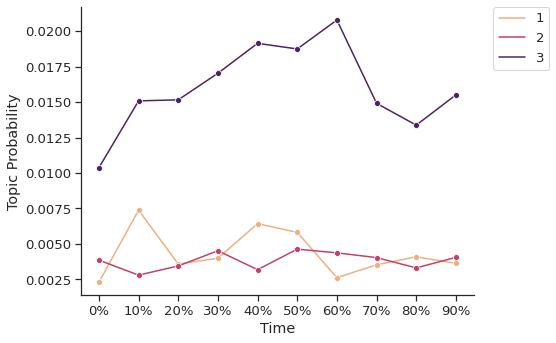}
    \caption[]%
    {{\small \textcolor{mygray}{T7: Virtual Interactions during COVID.}}}
\end{subfigure}
\hfill
\caption{\textcolor{mygray}{Time trends for topics focusing on virtual interactions and court proceedings.}}
\label{fig:time_plots_all_7_12}
\end{figure}

In addition, court proceedings are a significant part of the child-welfare process, and this topic emerges as significant at key decision points of the life of a case. As depicted in Figure \ref{fig:time_plots_all_7_12}(a), for G1 and G2, we observe upticks in trends towards the beginning of the case as well as a rise in trends towards the end. This matches our expectations since critical court hearings occur at the onset and towards the closing of a case for these groups. For G3, we observe several upticks in trends (spread out evenly) since the more severe cases of neglect/abuse require more court appearances in terms of reunification hearings, transfer of guardianship, or termination of parental rights. As depicted in Figure \ref{fig:time_plots_all_7_12}(b), we anticipate that the COVID pandemic may have also influenced the trends for these groups. During the pandemic, resources were directed towards cases that most needed them. Court hearings, parenting classes, and services were rescheduled and/or postponed for several cases in G1 and G2. Our collaborators at the agency shared that virtual court hearings, virtual classes, and virtual visitations were still being conducted for high needs cases, i.e. – most families in group G3.

\subsubsection{\textbf{Managing Medical Consent, Medication Administration, and Medical Appointments}}
\begin{figure}[H]
\vspace{-0.2cm}
\centering
\begin{subfigure}[b]{0.48\textwidth}
    \centering
    \includegraphics[width=\textwidth]{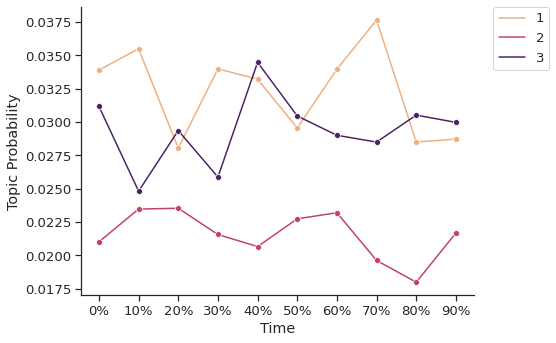}
    \caption[]%
    {{\small \textcolor{mygray}{T3: Managing Medical Consent}}}
\end{subfigure}
\hfill
\begin{subfigure}[b]{0.48\textwidth}  
    \centering 
    \includegraphics[width=\textwidth]{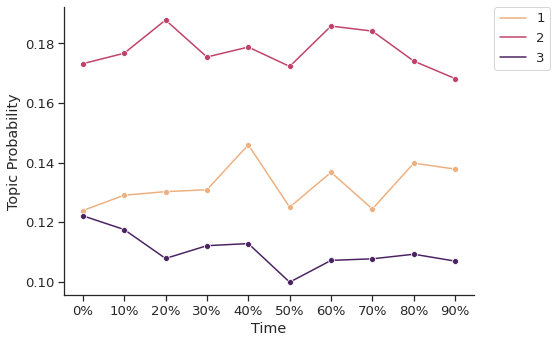}
    \caption[]%
    {{\small \textcolor{mygray}{T11: Managing Medication Schedules}}}
\end{subfigure}
\hfill
\vspace{-0.3cm}
\caption{\textcolor{mygray}{Time trends for topics focusing on medical consent and schedules management.}}
\vspace{-0.3cm}
\label{fig:time_plots_all_3_11}
\end{figure}

CW staff help manage medical consent between caregivers (topic 3) and help manage medication schedules (topic 11) for foster children. Topic 3 consistently emerges for G1 because CW staff discuss medical consent with birth parents early on and take children to necessary medical appointments (e.g., neglected dental health). This topic is also more significant for G3 (as compared to G2) because these are cases where more significant abuse/neglect may have occurred, and consequently, children are enrolled in services (e.g., individual therapy) to address their needs and the underlying trauma. We also anticipated this topic to emerge as more significant for G2 since medical consent needs to be managed between birth parents and foster parents and can lead to conflict. However, as depicted in Figure \ref{fig:time_plots_all_3_11}(b), managing medication schedules takes precedence for G2 because CW staff must continually ensure foster parents (especially short-term caregivers in G2) understand the medical needs of children and are giving them their medications per set schedule. This topic is less significant for G1 because most children are placed with birth parents and less significant for G3 since long-term caregivers are trained and certified in caring for high-needs foster children.

\subsubsection{\textbf{Coordinating Time, Travel, and Pickup Logistics for Visitations and Appointments}}
\begin{figure}[H]
\vspace{-0.1cm}
\centering
\begin{subfigure}[b]{0.48\textwidth}
    \centering
    \includegraphics[width=\textwidth]{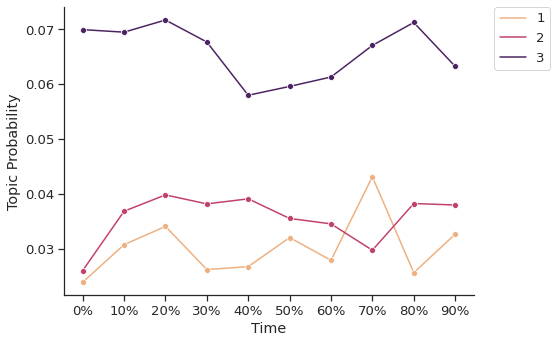}
    \vspace{-0.4cm}
    \caption[]%
    {{\small \textcolor{mygray}{T1: Scheduling Conflicts of Supervised Visitations}}}
\end{subfigure}
\hfill
\begin{subfigure}[b]{0.48\textwidth}  
    \centering 
    \includegraphics[width=\textwidth]{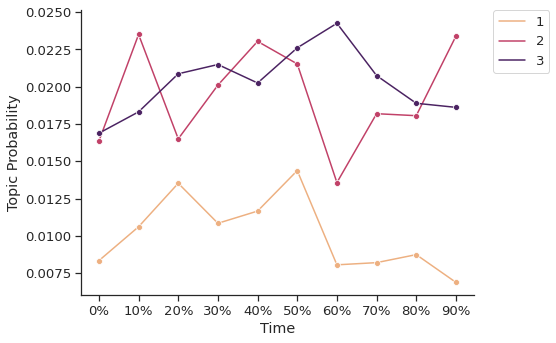}
    \vspace{-0.4cm}
    \caption[]%
    {{\small \textcolor{mygray}{T6: Managing time, travel, and pickup logistics}}}
\end{subfigure}
\hfill
\vspace{-0.3cm}
\caption{\textcolor{mygray}{Time trends for topics focusing on scheduling conflicts and time management.}}
\label{fig:time_plots_all_1_6}
\vspace{-0.5cm}
\end{figure}

Scheduling issues for supervised visits (topic 1) occur less frequently for groups G1 and G2; however, they are more common for G3 families. G3 includes cases of more severe neglect and/or abuse where intensive care is required in terms of medical appointments and supervised visits. For G3 cases, there may also be a no-contact order in place where parents can only see their children under proper supervision of family preservation caseworkers. However, as depicted by topic 6 (see Figure \ref{fig:time_plots_all_1_6}(b)), CW staff must also coordinate time, travel, and pickup logistics for court-ordered services, court hearings, visitations, and medical appointments. This topic emerges as significant for G2 at regular intervals since there may be multiple children involved in the case (and placed with different foster parents), and CW staff must coordinate these details among all parties. We observe two upticks in the trend for G1 and anticipate these to be medical appointments (general check-ups) conducted to assess children’s well-being before case closure.

\vspace{0.2cm}
\subsubsection{\textbf{Conducting Structured Assessments to Determine Risks and Progress}}
\begin{figure}[H]
\centering
\begin{subfigure}[b]{0.48\textwidth}
    \centering
    \includegraphics[width=\textwidth]{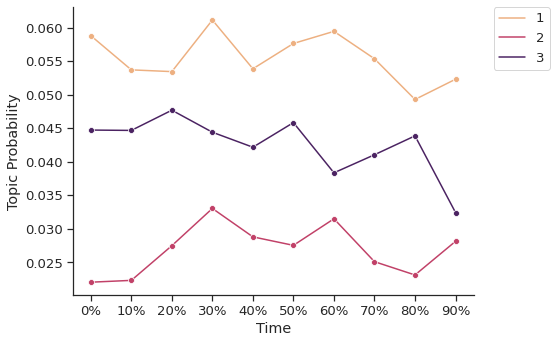}
    \caption[]%
    {{\small \textcolor{mygray}{T13: Progress in Parenting Classes}}}
\end{subfigure}
\hfill
\begin{subfigure}[b]{0.48\textwidth}  
    \centering 
    \includegraphics[width=\textwidth]{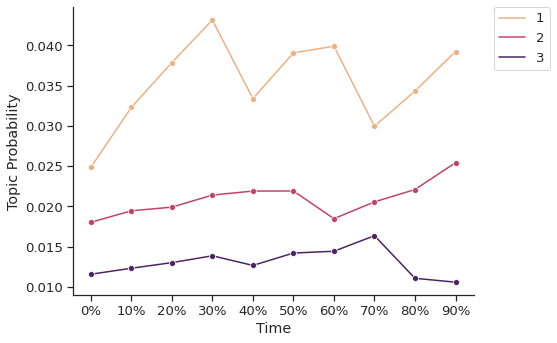}
    \caption[]%
    {{\small \textcolor{mygray}{T14: Safety Assessments during Home Visits}}}
\end{subfigure}
\hfill
\vspace{-0.2cm}
\caption{\textcolor{mygray}{Time trends for topics focusing on parenting classes and safety assessments.}}
\label{fig:time_plots_all_13_14}
\end{figure}

CW staff observe how parents respond to parenting classes and score their progress on quantitative structured assessments. This helps them assess the likelihood of the parents' employing these skills and strategies when addressing their children's needs and managing their behaviors. Topic 13 emerges consistently for both G1 and G3 (with upticks in trends spread out evenly) because parenting skills play an important role in achieving expedited reunification (as is the case with G1) but also in more severe cases of abuse/neglect as a means to assess if the parent is capable of meeting the needs of their children. We observe a similar trend for G2; however, the topic is less significant since more attention is being paid to managing logistics around multiple children, caregivers, and birth parents. CW staff also conduct home visits and score quantitative safety assessments from the onset of a case to assess if the home provides safe living conditions for children. Topic 14 emerges as being significant for G1, with several upticks in trends spread out evenly. For cases in G1, if the home provides a healthy and clean environment, then CW staff can provide in-home services to the families such that children's removal is not necessary. This topic did not emerge as significant for G2 and G3 (see Figure \ref{fig:time_plots_all_13_14}(b)) since the children are mostly placed with foster parents.

\subsection{Power analysis of personas}
Complex sociotechnical domains such as the child-welfare system consist of underlying power structures where some parties hold the majority of the power, exercise agency, and exert control over other parties. Power relationships with respect to CWS have been studied extensively in sociology literature \cite{johnson2018ambiguous, husby2018partnerships, reich2012fixing, bundy2009qualitative, rambajue2021intersectional}, however, computational text analysis of caseworkers' narratives to uncover such underlying power structures is an understudied topic.

\begin{figure}[!htb]
\centering
\begin{subfigure}{0.32\textwidth}
\centering 
\includegraphics[width=\textwidth]{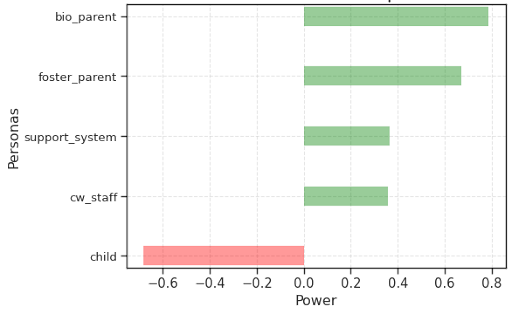}
\caption{\footnotesize{Group 1}}
\end{subfigure}
\begin{subfigure}{0.32\textwidth}
\centering 
\includegraphics[width=\textwidth]{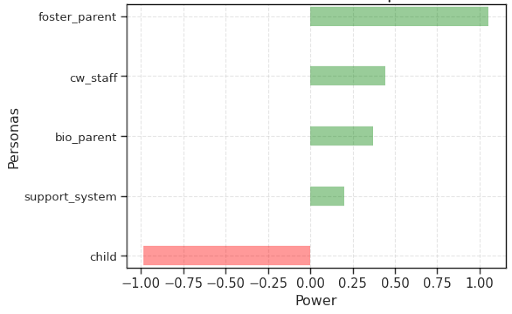}
\caption{\footnotesize{Group 2}}
\end{subfigure}
\begin{subfigure}{0.32\textwidth}
\centering 
\includegraphics[width=\textwidth]{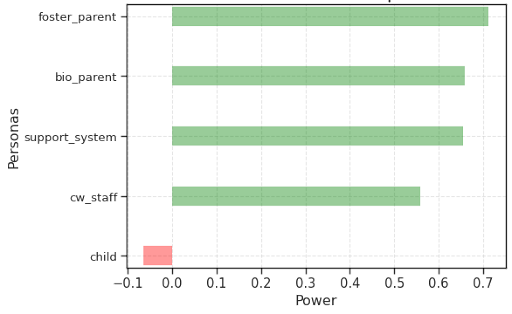}
\caption{\footnotesize{Group 3}}
\end{subfigure}
\vspace{-0.2cm}
\caption{Power scores for each persona across the 3 groups.}
\label{fig:power_scores}
\end{figure}

\begin{figure}[!htb]
\centering
\begin{subfigure}{0.32\textwidth}
\centering 
\includegraphics[width=\textwidth]{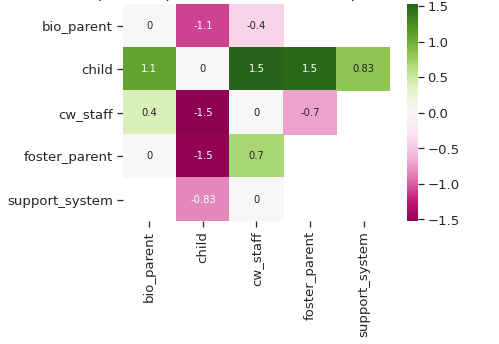}
\caption{\footnotesize{Group 1}}
\end{subfigure}
\begin{subfigure}{0.32\textwidth}
\centering 
\includegraphics[width=\textwidth]{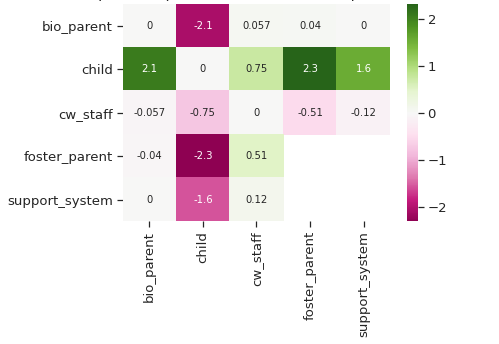}
\caption{\footnotesize{Group 2}}
\end{subfigure}
\begin{subfigure}{0.32\textwidth}
\centering 
\includegraphics[width=\textwidth]{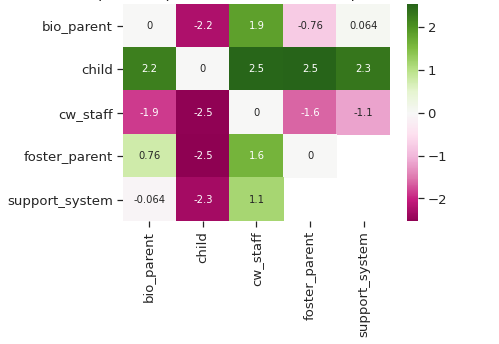}
\caption{\footnotesize{Group 3}}
\end{subfigure}
\vspace{-0.2cm}
\caption{Estimated power of personas (rows) on other personas (columns).}
\label{fig:power_effect}
\end{figure}

We conducted power analysis of casenotes and focused on five key personas which are actively involved at the front-end of child-welfare cases, namely, CW staff, birth parents, foster parents, birth parents' support system, and the foster child. Results of this analysis are depicted in Figure \ref{fig:power_scores} which shows power scores for each persona across the 3 groups, and Figure \ref{fig:power_effect} which demonstrates the estimated power of personas over other personas. Below, we first interpret our results for each of the three groups and then compare our findings across the three groups.

\textbf{Group 1} (1-10 interactions with CWS): As illustrated in Figure \ref{fig:power_scores} (a), birth parents have the most power for this group. As previously noted, cases in this group generally involve neglect (lack of childcare, lack of adequate food/clothing) and require birth parents to make adequate changes within their household to be able to provide a safe living environment for children. CW staff and the parents' support system are able to assist them but the parents must exercise their agency and demonstrate the necessary changes in their lives such that all agreed upon court conditions are met and CW staff can recommend reunification at the court hearing. Foster parents exhibit the most power after the birth parents since they actively care for a child daily and share the child's needs and behaviors with CW staff (which informs case planning and service delivery). Finally, as expected, foster children exhibit the least power (negative score) among all the personas.

\textbf{Group 2} (11-40 interactions with CWS): As illustrated in Figure \ref{fig:power_scores} (b), foster parents exhibit the most power for this group. Foster parents are the primary caretakers for this group and are actively involved in case planning. Prior studies conducted with CWS in Wisconsin showed that foster parents exercise the most agency with respect to how the needs and risks associated with foster children are assessed \cite{saxena2020child} and how much they are compensated by the state \cite{saxena2021framework2}. CW staff exhibit the most power after foster parents since they manage all the logistics associated with foster placements, such as finding resources for children, managing medical consent and medication schedules, scheduling visits, etc. The primary goal of CW staff is to ensure that foster parents are fully supported, and placement is not disrupted. Moving between different foster homes adversely affects foster children who develop emotional and behavioral problems and are unable to form meaningful relationships \cite{blakey2012review}. Birth parents exhibit lower power scores as compared to foster parents and CW staff because they may feel disempowered by the child-welfare process where their kids are removed and placed with multiple different foster parents. As previously noted, there may also be a lack of trust between birth parents and foster parents because of a lack of interpersonal relationships and ambiguity due to overlapping caregiving roles.

\textbf{Group 3} (40+ interactions with CWS): As illustrated in Figure \ref{fig:power_scores} (c), foster parents exhibit the most power for this group. Group 3 involves cases where severe abuse and/or neglect has occurred and requires trained and certified caregivers to meet these needs. There is a dearth of good foster homes in CWS where foster parents are trained in caring for high-needs kids \cite{curtis2020dimensions, clemens2017educational}, and therefore, CW staff must prioritize maintaining and supporting these placements. As previously noted, foster parents in this group may also be next of kin. For either case, there is a stronger interpersonal relationship between the foster parents and birth parents, which would explain birth parents exhibiting the most power after foster parents. The higher magnitude of power scores across personas also provides some evidence of an integrated approach towards family reunification adopted by CW staff where all personas are involved in child care and provide caregiver support to each other. Birth parents in this group must also complete mandatory court-ordered parenting classes and other services (domestic violence, drug, and alcohol abuse, etc.), and consequently, progress towards reunification is contingent upon them fulfilling these requirements. 

\textbf{Comparing across Groups}: CW staff act in a supporting role for groups G1 and G3 and exercise the least amount of agency (except for the child) compared to other personas. However, they take a lead role in G2 with respect to handling logistics and trying to address systemic barriers so that expedited reunification can be achieved for families. The agency has specialized meetings in place, called Permanency Consultations, designed to promote collaborative decision-making and expedite reunification \cite{saxena2021framework2}. As previously noted, if reunification does not occur within 15-months of a family being referred to CWS, the agency must begin exploring alternate placement options, that is, long-term foster care. CW staff's main objective is to prevent G2 cases from transitioning into G3 since long-term foster care leads to poor well-being outcomes for foster children. Moreover, finding good foster placements that can care for high-needs kids is hard because of a lack of good foster homes in the system. This is also why CW staff maintain a lower power profile with respect to foster parents. It is imperative that CW staff maintain good working relationships with both short-term and long-term foster parents so that there are homes to place children in need of care. Finally, CW staff (when acting in a supporting role for G1 and G3) also exercise less power as compared to birth parents' support system. They try to get the support system involved in the family's life such that birth parents have additional caregiver support and trusting relationships that they can rely on during times of crisis. This lowers the likelihood that the case would be re-referred to CWS due to instances of neglect (lack of childcare, lack of adequate food/clothing).

\textbf{Comparing across Personas}: Figure \ref{fig:power_effect} depicts a heatmap of power relationships between pairs of personas. As highlighted in prior work \cite{antoniak2019narrative}, it is possible for a persona to have a lower (or higher) cumulative power score but a higher (or lower) power score when only their interactions with another persona are measured. Interestingly, foster children who exhibit the least cumulative power appear to exercise more power over all individual personas. This could provide evidence for why CW staff work closely with birth parents in parenting classes so that parents are able to manage the behaviors of their children and regain agency in setting healthy disciplinary boundaries. Similarly, a successful foster placement requires that foster parents are able to manage the behaviors and needs of children. Inability to manage these needs/behaviors leads to placement disruptions where foster parents feel disempowered and put in their notice to end a placement; a significant ongoing concern in CWS \cite{carnochan2013achieving}. Surprisingly, CW staff appear to exercise the least amount of power (across all personas). Even for group G2, where CW staff assume a lead role, they appear to be sharing power across all individual personas.

\section{Discussion}

\subsection{Unpacking Invisible Patterns of Street-Level Discretionary Work (RQ1)}

Our topic modeling results highlight much of the hereto hidden, street-level discretionary work that caseworkers undertake while helping families (e.g., managing medication schedules, conducting quantitative assessments, establishing caregiving roles, navigating court proceedings etc.). These casenotes are collectively curated by CW staff involved at the front-end of case planning and offer a holistic picture for collaborative decision-making \cite{cpsmanual, saxena2021framework2}. What makes our results really important is that they revealed patterns of street-level work that were not even uncovered during an extensive ethnography at the same agency comprising of observations of collaborative meetings and interviews with caseworkers to understand their daily work practices and perspectives on algorithmic decision-making tools \cite{saxena2021framework2}. For instance, caseworkers help manage medical consent, medication administration (topics 3 and 11), as well as accompany clients to medical appointments and court hearings (topics 3, 11, 12). These topics were not highlighted during the ethnography even though they are collaboratively discussed in the casenotes. This suggests that qualitative deconstruction of work practices may not reveal all the nuances of invisible labor and in fact, demand complementary methodological lenses. By extension, we believe that this advocates for a need for both a qualitative and quantitative critique of sociotechnical systems. Critical computing has become popular at SIGCHI in recent years but remains bounded mostly by qualitative investigations \cite{CHI_SC, khovanskaya2020bottom, dombrowski2017low}. Moreover, we also examined the most recent job descriptions of the child-welfare caseworker positions \cite{ocm_childrens_2021, ocm_sainta_2021} at two CW agencies in the region and found that these patterns of work were not formally outlined in them either. The job descriptions state that caseworkers must complete documentation for court work as mandated by state law, but as revealed by topic 12 (and exemplar sentences), caseworkers are accompanying parents to court in order to assist them through the court proceedings. On the other hand, both job descriptions reveal that caseworkers must "conduct and document safety assessments." However, as illustrated by Saxena et al. \cite{saxena2021framework2}, data from these quantitative assessments are now being used to develop algorithmic risk assessments. As outlined by prior researchers \cite{zavrvsnik2019algorithmic, greene2020hidden, saxena2020human, saxena2021framework2}, these assessments and the administrative data used to build them are fundamentally biased. In contrast, our results point out that quantitative analysis of caseworker narratives can support strength-based, holistic assessments \cite{topitzes2019trauma} without being bogged down in the quagmire of biased algorithmic risk assessments. 

In sum, computational text analysis of casenotes helped uncover patterns of street-level discretion -ary work conducted by caseworkers that is otherwise hidden even from the findings of an ethnography or job descriptions of CW positions. This suggests two broader implications for SIGCHI research - a need for computational critique as well as a motivation to shift from biased risk assessments to more holistic strength-based assessments \cite{badillo2018chibest, zimmerman2013resiliency, topitzes2019trauma}.


\subsection{Understanding Constraints on Child-Welfare Practice (RQ2)}

Our results also highlight how constraints affect the work (discretionary or otherwise) that caseworkers need to do in order to provide better outcomes for children. We find that all children in CWS are not treated the same as some have higher needs than others (hence, our groups - G1, G2, G3). This differential need is affected by constraints (e.g., resource, bureaucratic, temporal, algorithmic, or other) and has been noted in prior work \cite{carnochan2013achieving, saxena2020human}. SIGCHI has become increasingly interested in the nature of work, especially when mediated/constrained by technology and algorithms \cite{holten2020shifting, ammitzboll2021street, de2020case}. As illustrated in our results in Section 5.2, different patterns (topics) of work are highlighted at different times through the life of a case and illustrate different interventions for different groups of families. For instance, as depicted in Figure 4(a), CW staff help secures essential resources for families. However, for G1 (less need), this generally takes the form of economic resources such as employment, food, clothing, and preventive services such as parenting classes. This requires CW staff to reach out to local parent support groups and family resource centers to connect clients to such services. Similarly, G2 (medium need) requires CW staff to find court-ordered services for their clients such as domestic violence classes, AODA (alcohol and other drug abuse) classes, therapy, etc. This requires CW staff to reach out to each of these service providers and find room for their clients. Much of this disparately available information can be curated into a system and made more accessible to CW staff. For instance, Yan et al. recently conducted an exploratory study to assess which systemic factors were associated with the services offered to clients. They offer direct implications for sociotechnical systems design in child welfare \cite{yan2021computational, buffalo_2021}. 

\vspace{0.1cm}
Similarly, as Figure 6(b) illustrates, CW staff spend a significant amount of time through the life of a case for Group 2 in making sure that foster parents are actively following the medication schedules for foster children. As previously noted (see Table 9), Group 2 generally involved multiple children placed with multiple different foster parents. CW staff must call foster parents (and do this for all their G2 cases) and make sure that the schedules are being followed. On the other hand, families in G3 have significant needs and require more care. Here, CW staff develop individualized trauma-responsive services (e.g., cognitive behavioral therapy, cognitive processing therapy etc.) for clients through specialized consultation sessions with medical professionals \cite{topitzes2019trauma}. As illustrated by Saxena et al. \cite{saxena2021framework2}, the CWS agency designed a holistic, strength-based algorithm centered in trauma-informed care to improve collaborative decision-making for high-needs families.

\vspace{0.1cm}
Summarizing all of these, an important implication arises for algorithms in CWS. Much of the current work has focused on (a) developing more sophisticated machine learning based risk assessment algorithms to improve the status-quo \cite{chouldechova2018case, de2020case} or (b) understanding breakpoints, biases, and ways in which caseworkers make decisions from currently implemented algorithmic constraints \cite{chor2013overview, chor2015out, saxena2020child, saxena2021framework2}. What is left unexplored at the current moment is whether (a) we need to be developing machine learning applications in CWS in the first place as well as (b) if simpler, non-algorithmic technological applications can help in removing some existing constraints that caseworkers work around (e.g., checking and notifying medication schedules). This aligns with recent work in worker-centered design in SIGCHI where caseworkers at a job placement center were averse to the introduction of a risk assessment algorithm (for profiling individuals) and instead, asked for sociotechnical systems that would help mitigate organizational constraints and help streamline bureaucratic processes \cite{holten2020shifting}. Caseworkers also perceived algorithms to be useful if they could support caseworkers' practices in strengthening cases that required additional resources \cite{ammitzboll2021street}. Similarly, caseworkers in child-welfare found utility in a simple decision-tools that help guide their decision-making processes through a trauma-informed care framework instead of predicting an outcome of interest \cite{saxena2021framework2}.


\subsection{Uncovering Latent Power Relationships in Child Welfare Systems (RQ3)}

\textit{Limitations.} We note important limitations of this computational power analysis approach that other researchers must consider before adopting this method. First, we acknowledge that this approach cannot uncover deep, structural power issues that are systemically embedded within CWS. We direct the readers to the works of J. Khadijah Abdurahman \cite{abdurahman2021calculating} and Victoria A. Copeland \cite{copeland2021s} who have deeply studied power hierarchies in CWS and illustrated that caseworkers both exercised power and experienced power asymmetries from supervisors, agency policies, and the court system. Second, since these casenotes are written by caseworkers, they do not capture families' firsthand accounts of their interactions with the system. We considered this methodology to be appropriate for this study for two reasons: 1) casenotes in our dataset are primarily written by the family preservation team whose goal is to achieve reunification for children and birth parents. That is, the team's objectives are aligned with those of the parents and centered in helping them prepare and achieve a favorable decision in court, and 2) casenotes are collaboratively written by case management and family preservation workers which adds a layer of accountability in regard to observations being recording in these casenotes. As illustrated by our recent study \cite{saxena2022chilbw}, this analysis would be inappropriate to study the casenotes of initial assessment/investigative caseworkers who exercised more power over families in regard to data being collected about parents and how critical decision were made. However, such quantitative analyses help illustrate these complexities within child welfare where different teams assume different roles.

\vspace{0.1cm}
We draw from existing SIGCHI scholarship on sociotechnical power rooted in feminist HCI \cite{bardzell2011towards} and worker-centered design \cite{fox2020worker, khovanskaya2021tools} to unpack our findings. From this theoretical scaffolding, we further distinguish between two kinds of power - first, the structural power that is systemic within any complex sociotechnical system as well as the power that exists as a result of day-to-day work relationships. We acknowledge that computational power analysis cannot structural power issues but rather surface the power complexities that arise from  daily work relationships. Further, we draw from Starhawk's \cite{starhawk1987truth} and Berger's \cite{berger2005power} disentanglement of these relationships between power-over, power-with, power-to, and power-within relationships. Largely, we find that in addition to the expected power-over relationships that one might expect to find within CWS stakeholders, there also exist some other kinds of unexpected power relationships that complicate some popular media narratives on CWS \cite{dettlaff2011disentangling, nourie2021child}. 

Our results find some evidence to support that CW staff exercised a more collaborative, power-with role (among adults) when they played a supporting role for groups G1 and G3 and only assumed more power-over relationships (in the case of group G2) when the primary goal was to expedite reunification such that cases did not transition into long-term foster care (i.e., group G3). This also provides some evidence for the efforts made within CWS from both a policy and practice standpoint to transition towards a "Families as Partners" model \cite{rauber2009courthouse} where parents are supposed to act as equal partners in the case planning process and have agency in the decision-making process. As previously noted, critical decision-making power in regard to reunification and termination of parental rights sits with the legal parties (i.e., - district attorneys, judges) \cite{carnochan2006child, ellis2010child} and often frustrates CW staff who are working with birth parents in their efforts to achieve reunification. These tensions between the court system and CW staff are well-documented in social work literature \cite{carnochan2006child, duffy2010macro, ellis2010child}. However, as previously noted, this is not to say that CWS is not riddled with deep systemic issues that disproportionately impact families of color \cite{johnson2009addressing, pon2011immediate}. On the contrary, our hope with this analysis is to illustrate the daily, working power complexities within this domain of which CW staff is only a piece of the complex puzzle comprising several parties with conflicting interests. For instance, a case typically involves four attorneys - one for each parent, the agency, and the child(ren) where each of these attorneys advocates for the individual rights of their clients \cite{vandervort2008legal}.   

Different power relationships also help uncover the differences in different families (i.e., - the three groups) involved in child-welfare and highlight the need to support both the families and CW staff in different capacities. For instance, CW staff is involved in a supporting, power-to relationship in both G1 (less need) and G3 (most need) groups, where they help secure resources for families. However, for G1, this translates into finding material resources (adequate food/clothing, childcare). Whereas, for G3, CW staff must find ongoing professional services (e.g., therapy domestic violence). On the other hand, G2 cases require that CW staff have a more power-over role in managing the needs of multiple foster placements. Moreover, different power relationships also directly impact how data is collected about children, how their needs are assessed, and have serious implications for algorithmic decision-making. For instance, our prior ethnographic study conducted at this agency \cite{saxena2021framework2} revealed that foster parents exercised significant control over how children's risks and needs were quantitatively scored, which impacted their compensation rates and the services offered to children \cite{saxena2021framework2}. This in turn leads to the manipulation of data and the algorithm such that foster parents received higher compensations. In prior work conducted in CWS \cite{lyons2014use}, these power imbalances also generated perverse incentive structures for algorithmic decision-making based on mental health needs. Medical professionals exercise more power than other involved parties in regard to the quantitative scoring of the needs of children. Consequently, they are paid when needs are detected and interventions offered. That is, there were clear professional and financial incentives that encouraged the detection of needs and led to the manipulation of the algorithm \cite{lyons2014use}. On the other hand, CW staff were trained to conduct mental health assessments; however, the detection of needs invariably led to more work on their part because it required them to find and secure services for children. That is, the short-term incentive for CW staff was to not detect needs so as to limit the amount of work \cite{lyons2014use}.

In sum, our analysis unpacks different kinds of work power relationships (e.g., power-over, power-to \cite{starhawk1987truth, berger2005power} etc.) between CWS stakeholders depending on the context and align well with prior social work literature on power relationships in CWS \cite{bundy2009qualitative}. These results imply that human-centered algorithm design in child welfare needs to understand and consider these power relationships to support the primary objective of providing positive outcomes for foster children.

\section{Limitations}
Our study only used casenotes from one CW agency in a US midwestern state, so our findings may not be generalizable to other states where different policies and regulations impact daily processes and decisions. Nevertheless, this study offers the methodology to perform computational narrative analysis in other CWS contexts and can help generate similar insights. Moreover, although all caseworkers are trained to record interactions and decisions in casenotes, their writing styles may vary. For instance, some caseworkers may not write details about characteristics captured in assessments (e.g., living conditions when scoring home-safety assessments). Moreover, it is imperative to note that casenotes may contain more contextual information, however, they are still based in workers' impression of family circumstances and could potentially introduce biases into decision-making \cite{saxena2022chilbw}. Lastly, our study only focused on one unsupervised ML method. While LDA is a powerful tool that has enabled us to member check our results with interpretations from CW stakeholders, it is important to explore and compare the efficiency of other models.

\vspace{-0.2cm}
\section{Future Work}
Abebe et al. \cite{abebe2020roles} recently outlined the roles of computing in social change and argued that computing serves as \textit{rebuttal} where it can help illuminate the boundaries of what is technologically feasible and acts as \textit{synecdoche} when it uncovers and makes long-standing social problems newly visible in public discourse. The purpose of this study was to assume these roles and highlight complexities within child-welfare (i.e., invisible labor, systemic constraints, power asymmetries) that are often overlooked by computing professionals who develop algorithmic systems. In addition, as highlighted by our recent study \cite{saxena2022chilbw}, quantitative de-construction of algorithms can further reveal power asymmetries, concealed biases, and data collection processes where investigative caseworkers exercised more power over families. That is, quantitative methods helped us uncover systemic issues and disparities that were not brought to light by a prior extensive ethnographic study \cite{saxena2021framework2}. This is primarily the case because practitioners in any underfunded environment have high workloads and do not have the time or resources to examine their own work practices. In sum, future studies on complex sociotechnical systems must employ both qualitative and quantitative methods to develop a deeper understanding as well as assess the role and scope of computing in solving systemic and societal problems. This mixed-methods approach has been further developed and formalized into Human-Centered Data Science \cite{aragon2022human}, an interdisciplinary field that draws from human-computer interaction, social science, statistics, and computational techniques.

\vspace{-0.2cm}
\section{Conclusion}
This study offers the first computational inspection of casenotes and introduces them to the SIGCHI community as a critical data source for studying complex sociotechnical systems. We applied topic modeling with LDA on collaboratively curated case narratives by CW staff. The casenotes are highly contextual for every family yet carry similarities concerning the processes families follow in child-welfare, including critical decisions made and personas involved at the front-end of case planning. Our results show that on-the-ground caseworkers engaged in several patterns of hidden labor that were not uncovered in prior ethnographic work or depicted in job descriptions. Analysis of different cases (based on the number of interactions) revealed that CW staff need to support families differently and further helped contextualize the meaning of topics. For instance, CW staff acquired different resources for G1 families (less need) vs. G3 families (high need). Finally, power analysis of casenotes revealed the power asymmetries within CWS that contest the dominant societal narrative that caseworkers exercise significant autonomy and are responsible for the removal of children. The power asymmetries have implications for algorithmic decision-making as these latent power structures directly impact generated algorithmic decisions.

\vspace{-0.2cm}
\begin{acks}
This research was supported by National Science Foundation grant CRII-1850517 and the University of Toronto’s Human-AI Interaction Summer Research School (https://www.thai-rs.com). Any opinion, findings, and conclusions or recommendations expressed in this material are those of the authors and do not necessarily reflect the views of our sponsors or community partners. We would like to thank our collaborators at the child-welfare agency, \textit{Wellpoint Care Network}, for sharing the casenotes dataset and providing valuable feedback on our data analysis process. We would also like to thank Maria Antoniak, David Mimno, and Karen Levy for sharing their computational text analysis code which provided us with the groundwork necessary for conducting our analysis. Finally, we are also thankful for our study participants as well as the anonymous reviewers whose suggestions and comments helped improve the quality of this manuscript.
\end{acks}

\bibliographystyle{ACM-Reference-Format}
\bibliography{bibliography}

\newpage
\appendix

\end{document}